\documentclass[fleqn,usenatbib]{mnras}

\usepackage{newtxtext,newtxmath}

\usepackage[T1]{fontenc}

\DeclareRobustCommand{\VAN}[3]{#2}
\let\VANthebibliography\thebibliography
\def\thebibliography{\DeclareRobustCommand{\VAN}[3]{##3}\VANthebibliography}

\usepackage{graphicx}	
\usepackage{amsmath}	
\usepackage{bm}
\usepackage{hyperref}
\usepackage[english]{babel}
\usepackage{tabu}
\usepackage{amsmath}
\usepackage[caption=false, position=t, singlelinecheck=off]{subfig}
\usepackage[utf8]{inputenc} 
\usepackage[T1]{fontenc} 
\usepackage{verbatim}
\usepackage{newtxtext,newtxmath}

\newcommand{\hans}[1]{\textcolor{red}{#1}} 

\graphicspath{{./}{figures/}}

\makeatletter
\newcommand*\bigcdot{\mathpalette\bigcdot@{.5}}
\newcommand*\bigcdot@[2]{\mathbin{\vcenter{\hbox{\scalebox{#2}{$\m@th#1\bullet$}}}}}
\makeatother

\title[Forming the Seeds of Giant Planets]{Filling in the Gaps: Can Gravitationally Unstable Discs Form the Seeds of Gas Giant Planets?}

\author[H. Baehr]{
Hans Baehr$^{1,2,3,4}$\thanks{E-mail: hans.baehr@uga.edu}
\\
$^{1}$Department of Physics and Astronomy, The University of Georgia, Athens, GA 30602, USA\\
$^{2}$Center for Simulational Physics, The University of Georgia, Athens, GA 30602, USA\\
$^{3}$Department of Physics and Astronomy, University of Nevada, Las Vegas, 4505 South Maryland Parkway, Las Vegas, NV 89154, USA\\
$^{4}$Nevada Center for Astrophysics, University of Nevada, Las Vegas, 4505 South Maryland Parkway, Las Vegas, NV 89154, USA\\
}

\date{Accepted XXX. Received YYY; in original form ZZZ}

\pubyear{2023}

\begin{document}
\label{firstpage}
\pagerange{\pageref{firstpage}--\pageref{lastpage}}
\maketitle

\begin{abstract}
Circumstellar discs likely have a short window when they are self-gravitating and prone to the effects of disc instability, but during this time the seeds of planet formation can be sown. It has long been argued that disc fragmentation can form large gas giant planets at wide orbital separations, but its place in the planet formation paradigm is hindered by a tendency to form especially large gas giants or brown dwarfs. We instead suggest that planet formation can occur early in massive discs, through the gravitational collapse of dust which can form the seeds of giant planets. This is different from the usual picture of self-gravitating discs, in which planet formation is considered through the gravitational collapse of the gas disc into a gas giant precursor. It is familiar in the sense that the core is formed first, and gas is accreted thereafter, as is the case in the core accretion scenario. However, by forming a $\sim 1 M_{\oplus}$ seed from the gravitational collapse of dust within a self-gravitating disc there exists the potential to overcome traditional growth barriers and form a planet within a few times $10^5$ years. The accretion of pebbles is most efficient with centimetre-sized dust, but the accretion of millimetre sizes can also result in formation within a Myr. Thus, if dust can grow to these sizes, planetary seeds formed within very young, massive discs could drastically reduce the timescale of planet formation and potentially explain the observed ring and gap structures in young discs.
\end{abstract}


\section{Introduction}
\label{sec:intro}

Observations of young discs show a wealth of complex structures in millimetre dust emission \citep{Andrews2018a,Long2018,Clarke2018}, scattered light off the gas disc surface \citep{Muto2012,Dong2018a} and molecular tracers \citep{Boehler2018,Oberg2021}. In particular, these very early discs have been imaged with significant ring and gap structure in continuum dust observations, with potential explanations ranging from magnetic fields \citep{Bethune2017,Hu2022}, radial thermal variations \citep{Schneider2018a,Ueda2021}, to hydrodynamic instabilities and vortices \citep{Bethune2016,Kuznetsova2022}, among other theories. A common suggestion is that planet formation and growth is ongoing, such that a planet can halt the inward migration of dust and open up a gap \citep{Zhang2018,Lodato2019,Pinilla2020}.

While planets have not been directly detected in the locations of the dust gaps of observed discs, their presence is potentially inferred by the distortion to the local Keplerian gas flow caused by a planet as seen through velocity channel maps of CO gas \citep{Pinte2016,Teague2018,Pinte2020}. If planets are indeed a cause of these structures and planet formation around young stars takes place much sooner than once thought, the age estimates of these discs would require that planet formation is very fast, even at distant orbital separations where many of these dust gaps are located.

\begin{figure*}
\centering
\includegraphics[width=0.49\textwidth]{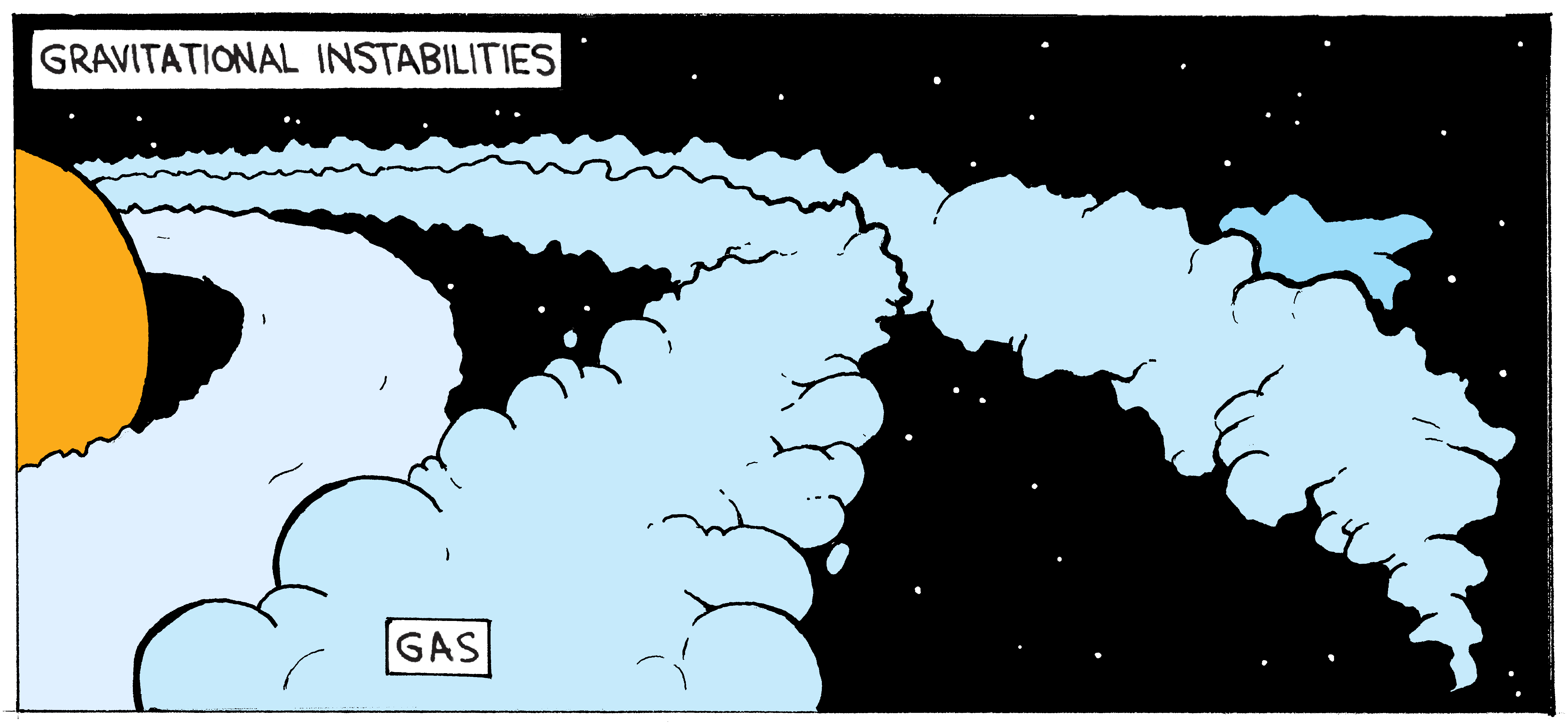}%
\includegraphics[width=0.49\textwidth]{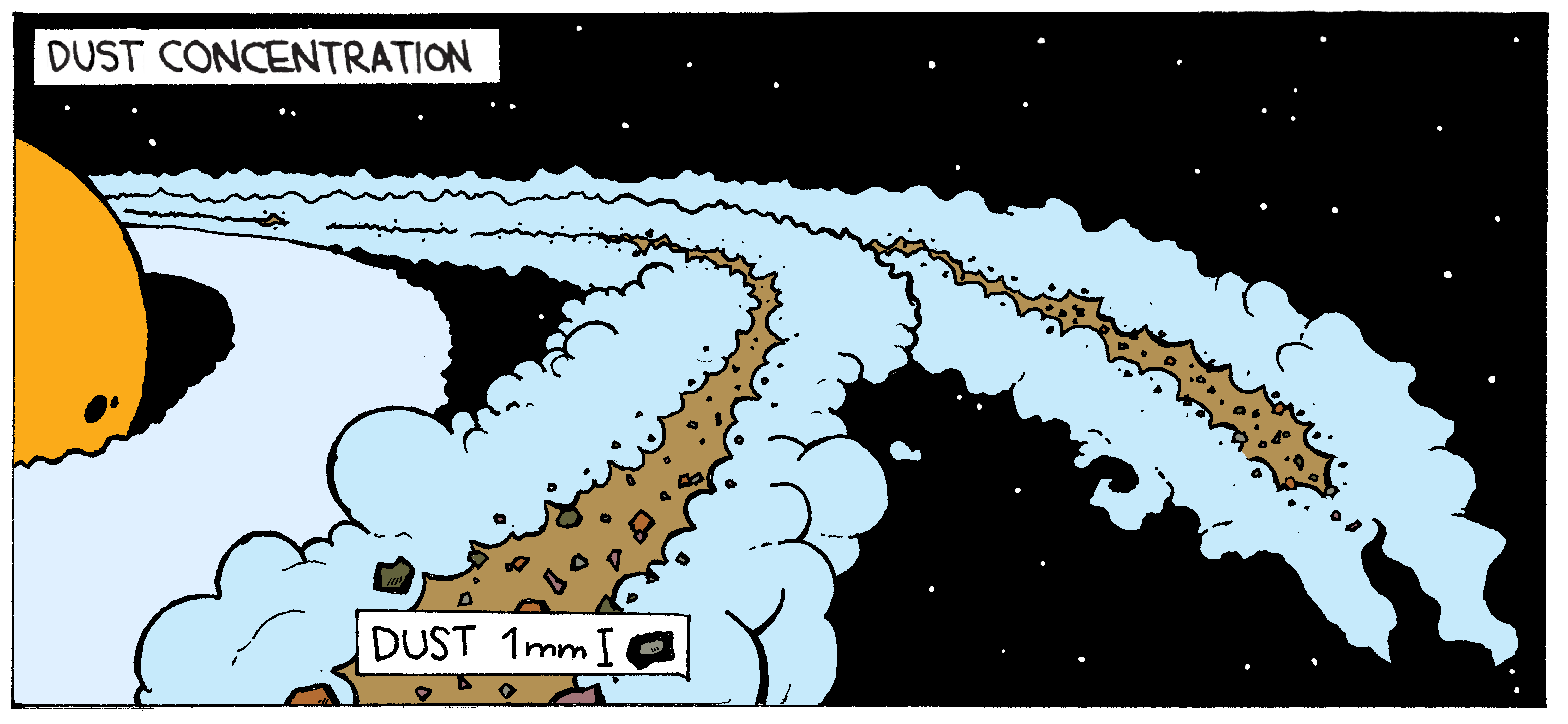}
\includegraphics[width=0.49\textwidth]{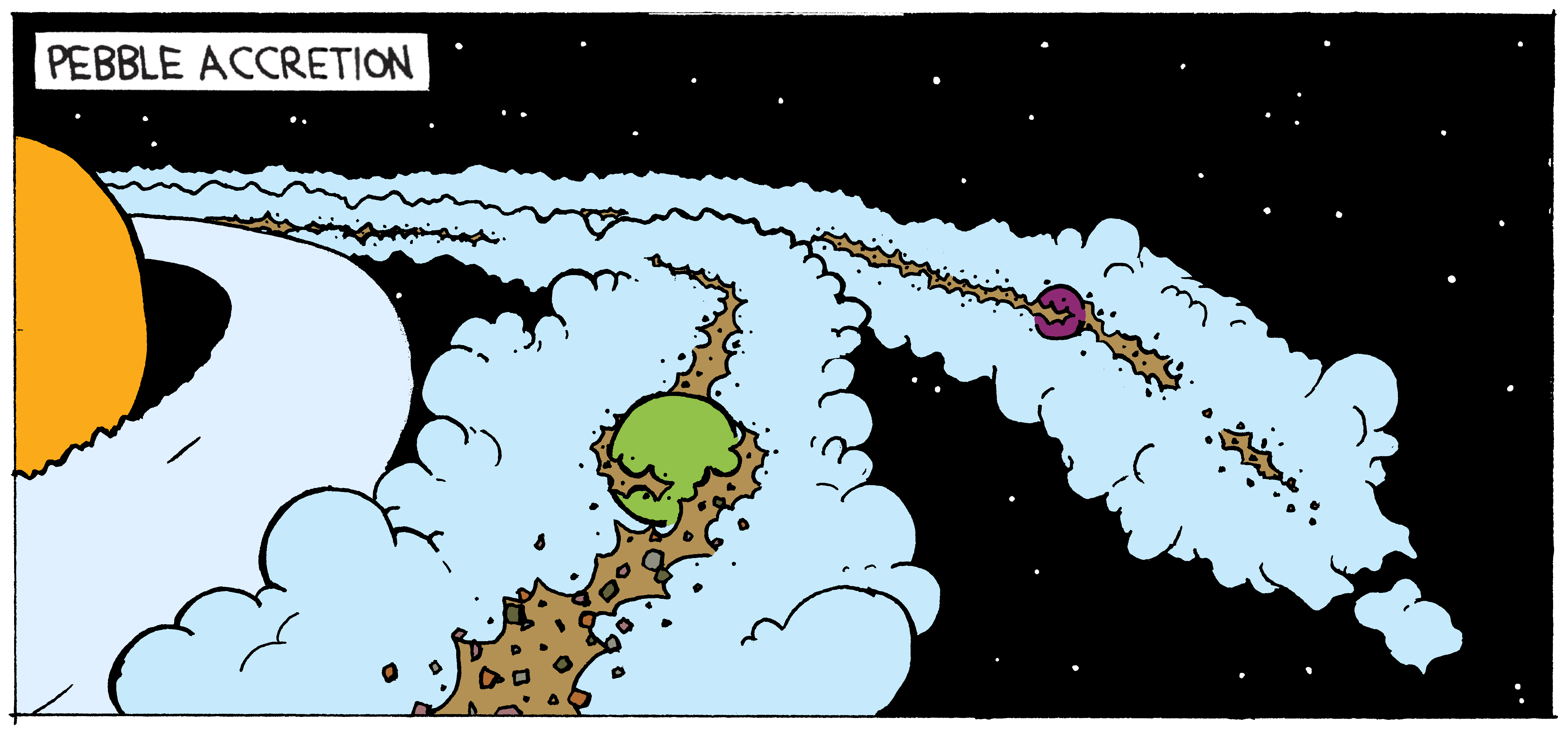}%
\includegraphics[width=0.49\textwidth]{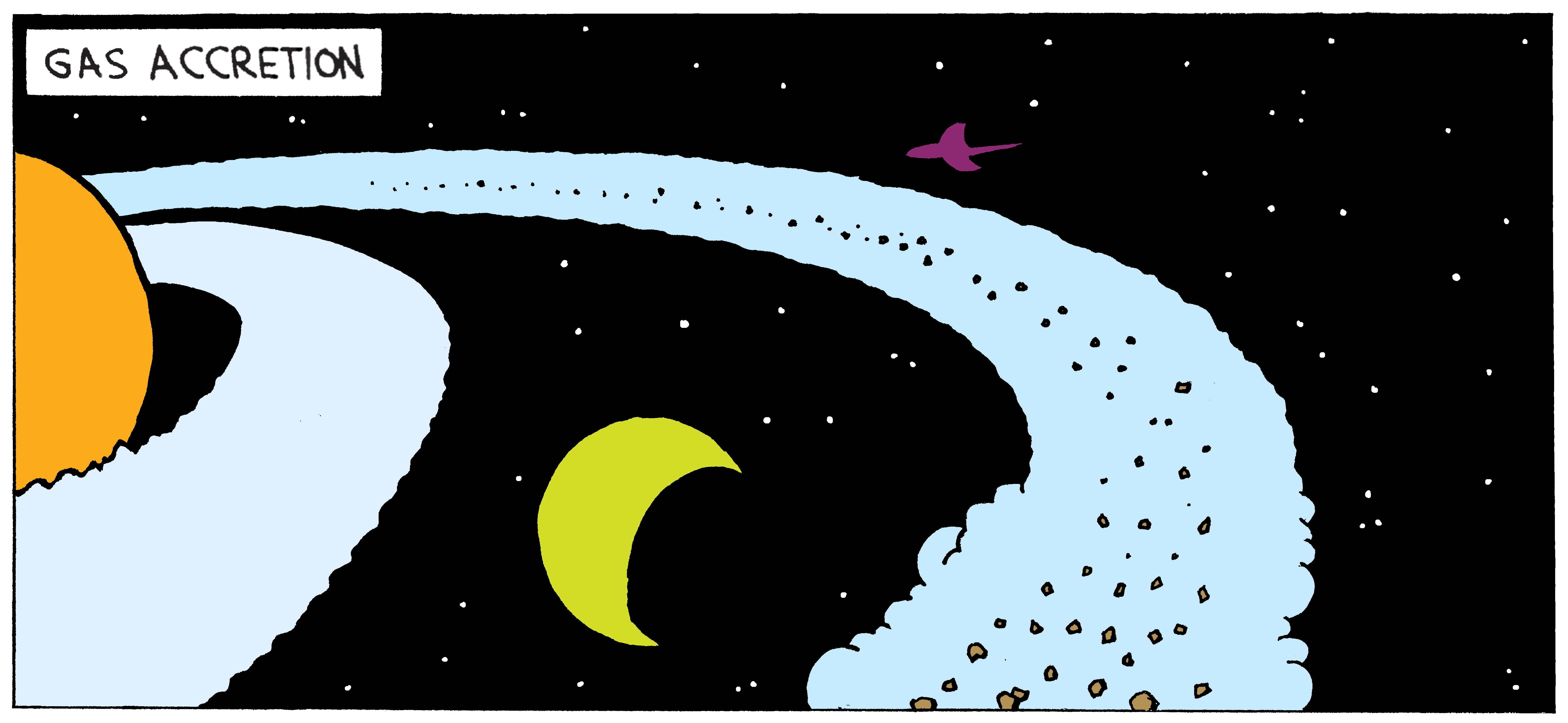}
\caption{A cartoon sketch of how the dust concentration in a self-gravitating disc can lead to rapid gas giant formation. Top left: a young, massive disc dominated by gas develops dense gas substructures; see Section \ref{sec:gravitoturbulentdiscs}. Top right: dust which has grown to millimetre sizes concentrates within the dense gas features, see Section \ref{subsec:graingrowth}. Bottom left: The gravitational collapse of the concentrated dust grains provides a seed for efficient pebble accretion which can rapidly grow to the size of a gas giant core, refer to Section \ref{subsec:dustcloudformation} and \ref{subsec:solidaccumulation}. Bottom right: With a large embedded core, gas will quickly fall onto the young planet and runaway accretion can clear a gap in the disc, see Section \ref{subsec:gasgiantgrowth}.}
\label{fig:schematic}
\end{figure*}

Disc fragmentation, where the gravitational collapse of the gas disc into a dense object, has been suggested as a possible pathway for the formation of gas giant planets on wide orbits \citep{Boss1997}. However, this idea has a number of disadvantages that make it difficult to reconcile with observed systems. First of all, disc fragmentation yields companion masses that are on the order of $10$ $M_{\mathrm{Jup}}$ at formation, which can grow substantially with later accretion from the disc \citep{Rafikov2005,Kratter2010,Rice2015,Kratter2016}. Secondly, unless fragmentation can concentrate significant amounts of dust at the dense centre immediately \citep{Baehr2019}, planets formed by disc instability are expected to have low metallicities and little or no solid core, which is not consistent with the gas giants in the solar system \citep{Bolton2017} or our understanding of gas giants in other systems \citep{Sato2005,Buhler2016}. Third, despite initially forming far out in the early disc, migration timescales are usually short enough that there is more than enough time for a gas giant planet formed through disc instability to become tidally disrupted or accrete onto the star \citep{Nayakshin2010,Baruteau2011,Zhu2012,Michael2012,Cloutier2013}. Finally, as it is currently understood, fragmentation requires that the disc be both especially cool and massive \citep{Cossins2009} and that the thermal relaxation (cooling) timescale is shorter than a few orbital timescales \citep{Gammie2001,Kratter2011}. Meeting both conditions in a realistic disc is possible but rare, considering that massive discs are typically optically thick with long cooling timescales, which is not favourable for disc fragmentation \citep{Haworth2020,Cadman2020}. However, the distant regions of the disc (beyond roughly 100 au) may become optically thin \citep{Rafikov2015}, in which case the cooling time can be short enough for fragmentation, provided densities are sufficiently high.

More familiar formation models, i.e. streaming instability or planetesimal and pebble accretion, are a possible way to form the large solid seeds of these gas giant planets, but may not be efficient enough in the outer disc where growth and collision timescales are longer \citep{Jang2022}. This difficulty can be inferred from semi-analytic planet formation models, which often require the existence of lunar-sized solid bodies to serve as the seeds of distant gas giant planets, when accounting for planet migration in a viscous disc \citep{Bitsch2015,Bitsch2019,Andama2022}. In population synthesis models, forming cold gas giant planets as observed in their present locations requires especially large discs that may be gravitationally unstable \citep{Schlecker2021}.

Instead of fragmentation of the gas disc, it has been argued that gravitoturbulent discs may produce an early population of planetesimals \citep{Rice2004,Gibbons2012,Gibbons2014a} and even planetary embryos \citep{Baehr2022}. The dust can concentrate within the dense gas structures prominent in marginally self-gravitating discs because they are efficient at concentrating dust through both vertical settling aided by the self-gravity of the disc gas \citep{Dipierro2015,Riols2020,Baehr2021a} and within the plane of the disc due to aerodynamic drag forces \citep{Baehr2021}. Planetary embryos formed in this way could then rapidly accrete solids and gas as in the conventional core accretion model and reach sub-Jovian to Jovian masses on timescales that are more compatible with observations. However, whether this scenario could realistically play a role in planet formation has not been fully explored. In the case that the disc soon becomes stable to gravitational perturbations, embryos deposited distances greater than 30 au could rapidly accrete gas and dust via pebble accretion \citep{Lambrechts2012,Morbidelli2012,Jang2022} such that a gap may be opened in the dust and/or gas component of the disc. The migration of a planet in such a scenario could be very rapid, leading to the loss of the planet to accretion onto the star. Thus, while calculating the growth timescales of seeds formed within a self-gravitating disc we will also compare them with typical migration timescales which consider the self-gravitating phase to be over.

In this paper we make the case for an alternate way in which self-gravitating discs can play a role in planet formation. We divide this analysis into three parts, starting in Section \ref{sec:gravitoturbulentdiscs} by setting up the overall physical picture of our self-gravitating disc assuming some period of steady gravitoturbulence. The second part in Section \ref{sec:formation} continues by constraining the valid dust and disc conditions where planetary embryos may be formed and the initial properties of these rocky bodies. Finally, in Section \ref{sec:evolution} we consider how these rocky bodies might evolve and grow in a still very gaseous disc which is well-supplied with pebbles for accretion. These stages are sketched out in Figure \ref{fig:schematic}. In the discussion, Section \ref{sec:discussion}, we address a number of consequences and alternatives as well as compare with observations of young discs with inferred or detected planets.

\section{Gravitationally Unstable discs}
\label{sec:gravitoturbulentdiscs}

In the aftermath of star formation, the newly forming disc around the young star accretes material from what remains of the natal molecular cloud, a.k.a. the stellar envelope. This supply of material can keep the disc mass at levels comparable to the protostar for a period of time whereupon the disc is unstable to gravitational collapse due to it's own self-gravity. The gravitational stability of massive discs to axisymmetric perturbations is described by the $Q$ parameter \citep{Safronov1960,Toomre1964,Goldreich1965}
\begin{equation} \label{eq:qparameter}
Q = \frac{c_s \Omega}{\pi G \Sigma}.
\end{equation}
A disc with local surface density $\Sigma$, Keplerian rotational frequency $\Omega$ and midplane sound speed $c_s$ can be said to be unstable to axisymmetric modes when $Q < 1$. For $1.4 \gtrsim Q > 1$, discs will still be unstable to higher order azimuthal modes which result in non-axisymmetric spiral density waves and possibly turbulence from gravitational stresses induced by these spirals. Discs need to be very cool or very massive to reach low values of $Q$, and thus gravitationally unstable discs are more likely beyond $\sim$ 30 au where accretion heating at the midplane is weaker and stellar irradiation dominates the vertical temperature structure.

We define radial scaling relations for a number of disc properties at a characteristic radius of $R = 50$ au using values from \citet{Baehr2022} for the midplane sound speed in the disc $c_s$:
\begin{equation} \label{eq:soundspeedprofile}
c_s = 199 \left( \frac{T}{11.25\,K} \right)^{1/2} \, m\, s^{-1}.
\end{equation}
We adopt a midplane temperature that scales with radius as $T = 11.25\, K (R/ 50\,au)^{-1/2}$ so the sound speed can be written as
\begin{equation} \label{eq:soundspeedradialprofile}
c_s = 199 \left( \frac{R}{50\,au} \right)^{-1/4} \, m\, s^{-1},
\end{equation}
and the orbital frequency $\Omega$ is
\begin{equation}
\Omega = 1.78 \times 10^{-2} \left( \frac{R}{50\, au} \right)^{-3/2} \left( \frac{M_{*}}{1 M_{\odot}} \right)^{1/2}\, yr^{-1}.
\end{equation}
From the above definitions, we can derive the scale height of the gas disc $H_{g} = c_s/\Omega$
\begin{equation}
H_{g} = 2.36 \left( \frac{T}{11.25\,K} \right)^{1/2} \left( \frac{R}{50\, au} \right)^{3/2} \left( \frac{M_{*}}{1 M_{\odot}} \right)^{-1/2}\, au,
\end{equation}
the disc aspect ratio
\begin{equation} \label{eq:discaspectratio}
\frac{H_{g}}{R} = 0.047 \left( \frac{T}{11.25\,K} \right)^{1/2} \left( \frac{R}{50\, au} \right)^{1/2} \left( \frac{M_{*}}{1 M_{\odot}} \right)^{-1/2}.
\end{equation}
and the orbital period $P = 2\pi\Omega^{-1}$
\begin{equation}
P = 353 \left( \frac{R}{50\, au} \right)^{3/2} \left( \frac{M_{*}}{1 M_{\odot}} \right)^{-1/2}\, yr.
\end{equation}
We initially adopt a gas surface density $\Sigma$ such that $Q=1$ at 50 au
\begin{equation}
\Sigma_{g} = 53 \left( \frac{R}{50\, au} \right)^{-3/2} \left( \frac{M_{*}}{1 M_{\odot}} \right)^{1/2} \,g\,cm^{-2}.
\end{equation}
From this point on, unless otherwise noted, all scaling relations will assume that $M_{*} = M_{\odot}$. A disc with this surface density profile has a monotonically decreasing value of $Q$ for increasing radius from the star, and is unrealistic when observed discs are truncated outside of a few hundred au \citep{Andrews2009}.  Thus, we add an outer exponential cutoff at a radius of $R_{\mathrm{cut}}$ to the power-law radial surface density profile

\begin{figure}
\centering
\includegraphics[width=0.48\textwidth]{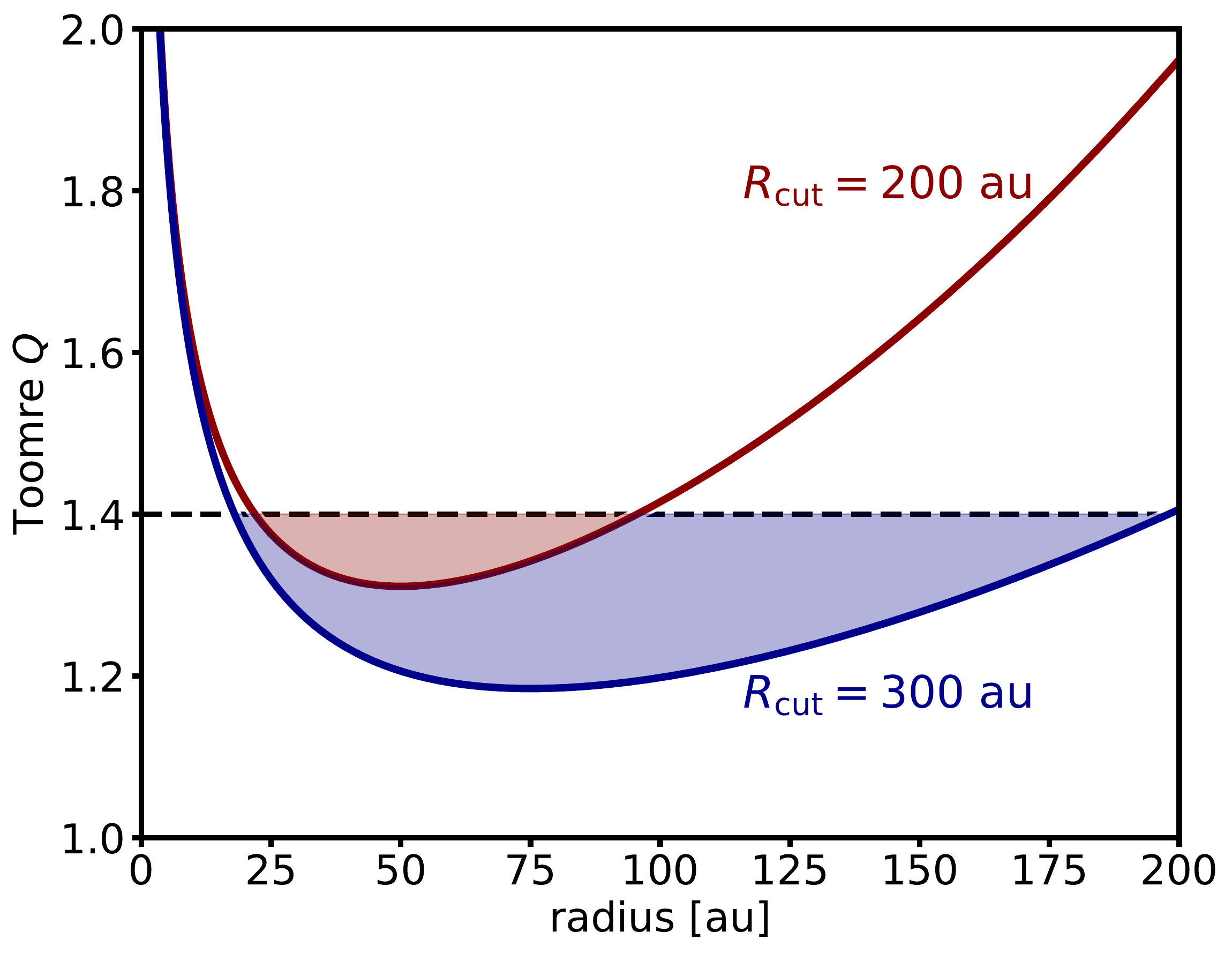}
\caption{An example of the disc stability $Q$ as function of radius as is considered in this analysis. A disc with a power-law surface density profile will become more unstable with increasing radius indefinitely, but a more realistic disc with an outer truncation, modeled in Equation \eqref{eq:toomrecutoff} with an exponential cutoff, will have a limited radial range of gravitational instabilities, which will vary depending on the cutoff radius $R_{\mathrm{cut}}$. The shaded regions indicate the radial locations of the disc where $Q < 1.4$ and non-axisymmetric (spiral) gas structures may form, which leads to dust concentration.}
\label{fig:radialprofiles}
\end{figure}
\begin{equation}
\Sigma_{g} = 53 \left( \frac{R}{50\,au} \right)^{-3/2} \exp \left( -\frac{R}{R_{\mathrm{cut}}}  \right) \,g\,cm^{-2}.
\end{equation}
From these definitions we derive a basic power-law $Q$ profile 
\begin{equation}
Q = \left( \frac{R}{50\,au} \right)^{-1/4},
\end{equation}
which we then modify with an outer exponential cutoff for a more realistic radial profile
\begin{equation} \label{eq:toomrecutoff}
Q_{\mathrm{cut}} = 1.21 \left( \frac{R}{50\,au} \right)^{-1/4} \exp \left( \frac{R}{R_{\mathrm{cut}}} \right),
\end{equation}
which is plotted for exponential cutoff radius $R_{\mathrm{cut}} = 200$ au and $R_{\mathrm{cut}} = 300$ au in Figure \ref{fig:radialprofiles} to show how the radial range of non-axisymmetric structure can change with the disc size. For the remainder of the paper, the default value will be $R_{\mathrm{cut}} = 200$ au unless otherwise stated. The profile as described in Equation \ref{eq:toomrecutoff} and shown in Figure \ref{fig:radialprofiles} intentionally assumes the disc to be quasi-steady and gravitoturbulent, because dust can still concentrate and gravitationally collapse even if the gas is marginally stable \citep{Longarini2023}.

The turbulent strength of a gravitoturbulent disc can be well described by the relation with the gas cooling timescale \citep{Gammie2001}
\begin{equation}
\alpha_{\mathrm{GI}} = \frac{4}{9}\frac{1}{\gamma(\gamma - 1)t_{c}\Omega}
\end{equation}
which for a cooling timescale that avoids fragmentation $t_{c} = 10\,\Omega^{-1}$ and an adiabatic index of $\gamma = 5/3$, yields a value of $\alpha_{\mathrm{GI}} \approx 0.01$. We will use this value as the starting point in our analysis, noting that the value could be very time and location dependent in a realistic disc. The disc profile is constructed such that non-axisymmetric density structures are the dominant feature and fragmentation is not expected due to the choice of cooling timescale greater than $\sim 5\Omega^{-1}$. The cooling timescale will slightly affect the strength of the spiral density perturbations $\delta\Sigma_{g}/\Sigma_{g,0} \propto (t_{c}\Omega)^{-1/2}$ \citep{Cossins2009}, which will impact how easily dust can concentrate.

Based on these estimates for a radial disc profile, we expect some region beyond $\sim$25 au to be moderately gravitationally unstable. Such a picture has generally been the theoretical expectation of massive discs \citep{Rafikov2015,Rice2015} and has recently been borne out in radial surface density estimates of discs like HL Tau \citep{Kwon2011,Booth2020}, GM Aurigae \citep{Schwarz2021} and AS 209 \citep{Powell2019,Franceschi2022}. The extent of the self-gravitating region will depend on the outer cutoff of the disc, as shown in Figure \ref{fig:radialprofiles}.

\section{Formation of Planetary Embryos}
\label{sec:formation}

With the preceding estimates for the radial profile of the gas disc, we turn to the dust. First, we need to consider the growth of dust from the micron-sized grains that dominate the interstellar medium (ISM) to the millimetre to centimetre sizes which are most conducive to concentration. After that we consider under what disc conditions can dust gravitationally collapse into bound clouds.

\begin{figure}
\centering
\includegraphics[width=0.48\textwidth]{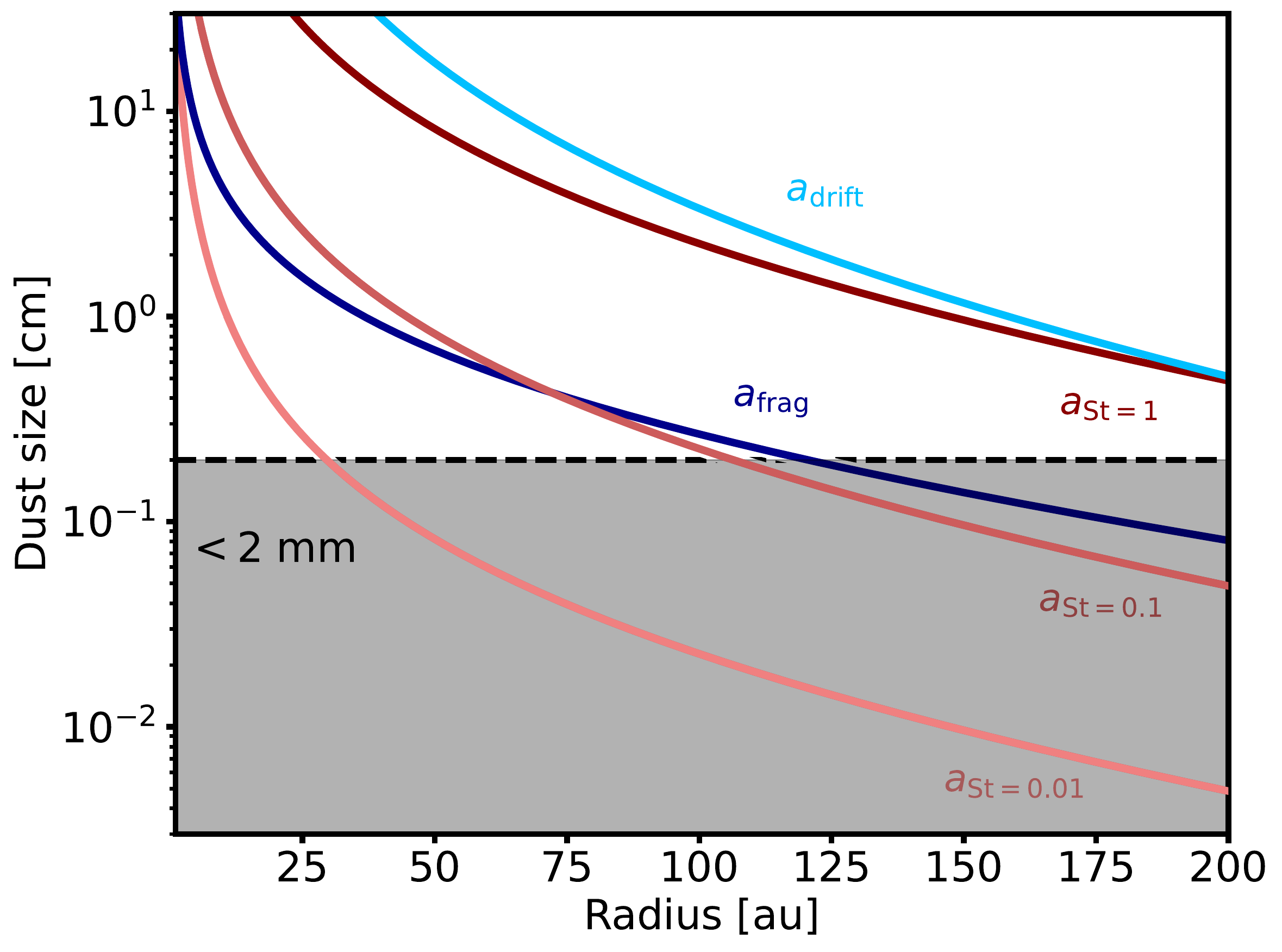}
\caption{Dust size as a function of radius for each dimensionless $\mathrm{St}$. Dust growth up to several centimetres is unlikely and thus $\mathrm{St}=1$ is not a realistic expectation at any distance from the star. A size of $\mathrm{St}=0.01$ corresponds to a size of roughly a millimetre and smaller outside of 30 au, which is a reasonable size for grains to reach through collisional growth. We compare with the drift and fragmentation limits for the dust to show that these limits prevent growth to the largest size under consideration $\mathrm{St}=1$ and millimetre sized dust beyond 150 au.}
\label{fig:radialdustsize}
\end{figure}

In the Epstein drag regime, particles are smaller than the mean free path of the gas and interact with the gas through random collisions with gas particles. As such, particle size can be expressed in terms of a dimensionless friction time $\mathrm{St}$ from the stopping time $t_s$:
\begin{equation} \label{eq:dustsize}
\mathrm{St} = t_s\Omega = \frac{a\rho_{\bigcdot}}{c_s \rho_{\mathrm{g}}}\Omega,
\end{equation}
where $a$ is the radius of a dust grain in centimetres and $\rho_{\bigcdot}$ is the material density of that dust grain, assumed equal to $2\,g\,cm^{-3}$. We then calculate the physical dust size as function of radius rewriting Equation \eqref{eq:dustsize} in terms of the gas surface density $\rho_{\mathrm{g}} = \Sigma_{\mathrm{g}}/\sqrt{2\pi} H_{\mathrm{g}}$
\begin{align}
a &= \frac{\mathrm{St}\, \Sigma_{\mathrm{g}}}{\sqrt{2\pi}\rho_{\bigcdot}}\\
&= 0.11 \left( \frac{\mathrm{St}}{0.01} \right) \left( \frac{R}{50\, au} \right)^{-3/2} \exp \left( -\frac{R}{R_{\mathrm{cut}}} \right)\, cm.
\end{align}
We plot this physical dust size as a function of radius for the three dimensionless dust sizes in Figure \ref{fig:radialdustsize}. As a first approximation, disc metallicity is assumed to be at standard interstellar medium (ISM) level abundances $Z=0.01$, but some amount of enrichment of the solid-to-gas ratio might be expected, either due to some natural variance or some large scale depletion of gas, i.e. photoevaporation or disc winds.

To get a sense for the importance of dust size, we compare with two common limitations to dust growth, the drift barrier and the fragmentation barrier. The limiting size of radially drifting dust particles is \citep{Birnstiel2012}
\begin{equation}
\mathrm{St}_{\mathrm{drift}} = Z \left( \frac{v_{k}}{c_{s}} \right)^{2} \left| \frac{\partial \ln p}{\partial \ln R} \right|^{-1}
\end{equation}
where $v_k = \Omega R$ is the Keplerian velocity and $p$ is the midplane gas pressure. For a midplane pressure with an isothermal equation of state $p = \rho c_s^2$, the radial pressure gradient term equals $2/7$. Written in terms of the physical size of the dust grain $a$:
\begin{equation}
a_{\mathrm{drift}} = \frac{\Sigma_{d}}{\sqrt{2\pi} \rho_{\bigcdot}} \left( \frac{v_{k}}{c_{s}} \right)^{2} \left| \frac{\partial \ln p}{\partial \ln R} \right|^{-1}
\end{equation}
When the dust size is larger than $a_{\mathrm{drift}}$, the inwards drift can outpace the growth to planetesimal size and places an upper limit on dust size for a given disc radius. The second limit to the dust size is the fragmentation barrier, such that above a certain size $\mathrm{St}_{\mathrm{frag}}$, the collisions between solids will reduce the size of the impactors, rather than lead to growth. This size limit can be expressed in terms of the dimensionless size
\begin{equation}
\mathrm{St}_{\mathrm{frag}} = \frac{1}{3}\frac{u_{f}^{2}}{\alpha c_{s}^{2}}
\end{equation}
or in terms of the physical size
\begin{equation}
a_{\mathrm{frag}} = \left( \frac{1}{18\pi} \right)^{1/2} \frac{\Sigma_{g}}{\rho_{\bigcdot}\alpha}\frac{u_{f}^{2}}{ c_{s}^{2}}
\end{equation}
Both of these size constraints are plotted in Figure \ref{fig:radialdustsize} assuming a fragmentation velocity of $u_{f} = 10\, m\,s^{-1}$, which is not uncommon for icy grains \citep{Wada2009,Gundlach2015} and turbulent $\alpha = 0.01$. Due to the strong turbulence associated with GI unstable discs, the fragmentation barrier is the stricter limit to dust growth in the disc, particularly in the outer disc where even millimetre grains are above the fragmentation limit.

\subsection{Growth of Dust Grains}
\label{subsec:graingrowth}

In the immediate aftermath of the collapse of the molecular cloud, the distribution of dust sizes will resemble that of the ISM \citep{Bate2017}, which is dominated by dust grains smaller than $\sim 1\,\mu$m and the abundance of larger grains decreases with a power law of index $n = -3.5$ \citep{Mathis1977}. The coagulation of smaller grains into larger ones through random collisions will flatten the distribution over time, a process that should be relatively efficient if early planet formation is to occur. 

The size of the dust disc is comparable to the size of the gas disc, but usually slightly smaller \citep{Hendler2020}. However, simulations of star and disc formation indicate that circumstellar discs are initially small \citep{Vorobyov2009a,Kuffmeier2017,Xu2021}, which may suggest that continued accretion from the envelope is necessary for gravitationally unstable discs to exist on the scale of hundreds of au \citep{Wang2023a}. The relative velocities of dust grains in gravitoturbulent discs can become destructive, particularly when as the grains become larger \citep{Booth2016}. This can potentially be mitigated by considering that dust grains in the outer disc where GI is active are likely icy, which can increase the threshold fragmentation velocity of grains \citep{Windmark2012}.

Turbulent discs can provide the necessary velocity fluctuations to facilitate collisions among small grains up to millimetre sizes, perhaps even in the presence of the strong turbulence of a self-gravitating disc \citep{Booth2016,Sengupta2019}. The timescale for settled dust to double in size through collisional interactions is \citep{Birnstiel2012}
\begin{equation} \label{eq:growthtimescale}
\tau_{\mathrm{grow}} \sim \frac{1}{Z} \Omega^{-1},
\end{equation}
but collisions alone are likely insufficient to explain the growth from micron-sized dust to millimetre grains in a young, embedded disc \citep{Tu2022}. Even so, we estimate the time to grow from micron to millimetre sizes as \citep{Birnstiel2012}
\begin{equation} \label{eq:growthtime}
t_{\mathrm{grow}} \approx \tau_{\mathrm{grow}} \ln \left( \frac{a_{1}}{a_{0}} \right) = \frac{9.2}{Z}\Omega^{-1},
\end{equation}
where $a_{1}$ is one millimetre and $a_{0}$ is one micron. Within 200 au this timescale is less than $5 \times 10^{5}$ years, which we will use as a conservative estimate and constraint on the available time for planet formation. In a gravitoturbulent disc, small grains will not be well settled and grain growth through collisions can be accelerated by turbulent stirring as well as by a sweep-up effect of larger grains \citep{Xu2023}

Observations of young class 0/I discs show dust can already grow up to millimetre sizes \citep{Sadavoy2019}, but dust growth up to several centimetres takes longer and unlikely in the quantity necessary for gravitational collapse and thus $\mathrm{St}=1$ is not a realistic target at any distance from the star. Dimensionless sizes $\mathrm{St}=0.1$ can be equivalent to a dust grain of a few millimetres, but only in the most distant regions of the disc. The growth of dust is most efficient in the inner disc unless turbulence can increase the rate of collisions in the outer disc \citep{Lebreuilly2021,Bate2022}, but it may be possible that jets around the protostar can extract inward drifting dust and redistribute them to the outer disc \citep{Tsukamoto2021}.

\begin{figure}
\centering
\includegraphics[width=0.48\textwidth]{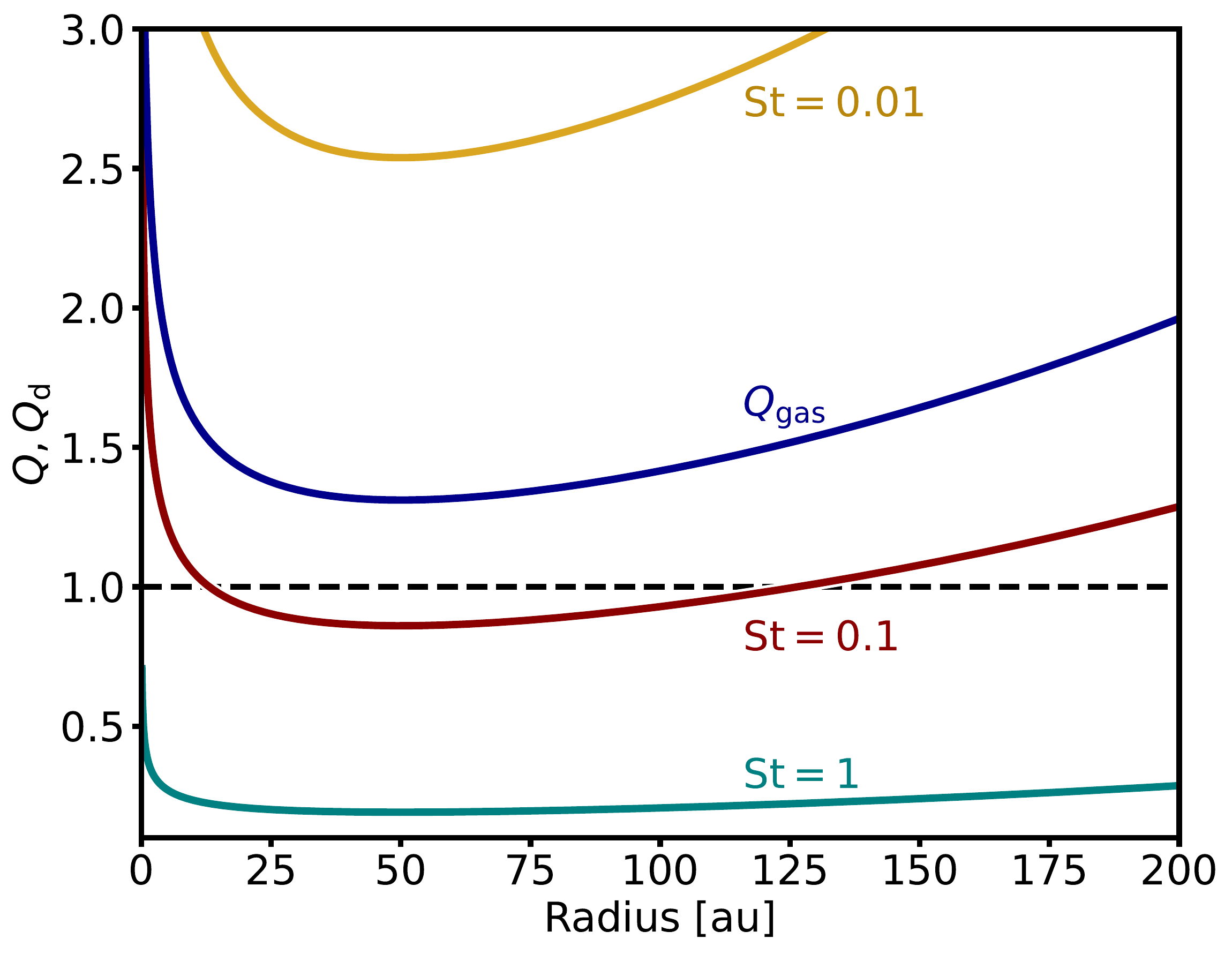}
\caption{The gravitational stability $Q_{\mathrm{d}}$ of three dust size species as a function of radius, assuming the gas disc has an exponential cutoff with $R_{\mathrm{cut}} = 200$ au. The value for the dust diffusion $\delta$ is taken from the simulations of \citet{Baehr2022} for $\mathrm{St}=1$ and $\mathrm{St}=0.1$, while for $\mathrm{St}=0.01$ the diffusion values come from the simulations of \citet{Baehr2021a}.}
\label{fig:radialdustprofiles}
\end{figure}

\begin{figure}
\centering
\includegraphics[width=0.48\textwidth]{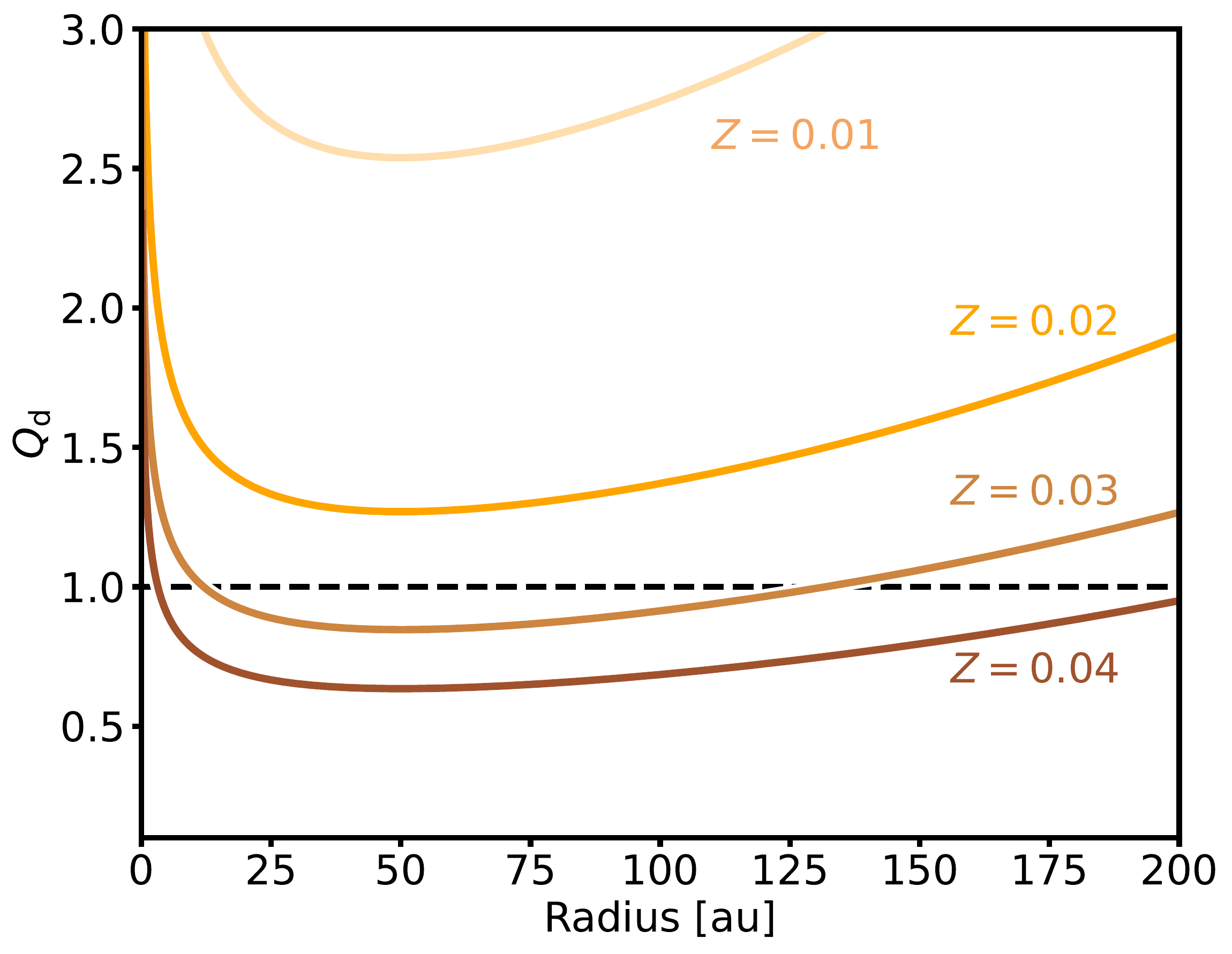}
\caption{Same as Figure \ref{fig:radialdustprofiles} except focusing only on dust of size $\mathrm{St}=0.01$ for different initial metallicities $Z$. Simply increasing the metallicity decreases $Q_{\mathrm{d}}$, near the unstable regime, but dust concentration $\epsilon$ is possibly increased as well (not included here), improving the chance of gravitationally collapsing dust.}
\label{fig:radialsmalldustprofiles}
\end{figure}

\begin{table}
	\centering
	\caption{Diffusion and concentration parameters used in Equation \ref{eq:gerbigparameter}.}
	\label{tab:parameters}
	\begin{tabular}{ccc}
		\hline
		Particle size $\mathrm{St} $ & diffusion $\delta$ & concentration $\epsilon$ \\
		\hline
		1 & $2.1\times 10^{-2}$ & 201 \\
		0.1 & $5.6\times 10^{-3}$ & 90 \\
		0.01 & $8.5\times 10^{-3}$ & 71 \\
		\hline
	\end{tabular}
\end{table}

Further evidence for grain growth in young discs can come from the spectral index \citep{Kwon2009,Harsono2018,Paneque-Carreno2021} or from models of chemistry and radiative transport \citep{Flores-Rivera2021}, which shows that the dust distribution of the ISM becomes flatter, particularly in the inner regions of the disc. Combined with polarimetry that can provide additional evidence for grain growth as well as information on the spatial distribution of larger grains \citep{Kataoka2015,Pohl2016a,Bertrang2017}, it becomes apparent that the first million years of disc evolution can lead to the growth necessary for gravitational collapse into bound clouds of dust.

\subsection{Gravitational Collapse to Bound Dust Clouds}
\label{subsec:dustcloudformation}

Provided dust grains in a cloud can reach a local density such that the particles cannot be sheared apart by the disc, the Hill density
\begin{equation}
\rho_{\mathrm{Hill}} = \frac{9}{4\pi}\frac{M_{*}}{R^{3}}
\end{equation}
the dust may collapse under it's own self-gravity. However, in addition, dust in a turbulent medium has an internal diffusion $\delta$ resulting from dust-gas interactions that can resist the self-gravity of dust \citep{Klahr2020}. We will define the local dust concentration of a cloud by the factor $\epsilon = \Sigma_{d}/\Sigma_{d,0}$. As a local pressure maxima, dense spiral waves of a self-gravitating disc are excellent locations to concentrate dust in the plane of the disc \citep{Baehr2021}. Considering an initial dust-to-gas mass ratio $Z = 0.01$, we define a gravitational collapse criterion for the dust as defined in \citet{Gerbig2020}
\begin{equation} \label{eq:gerbigparameter}
Q_{\mathrm{d}} = \frac{3}{2}\frac{Q}{\epsilon Z}\sqrt{\frac{\delta}{\mathrm{St}}}.
\end{equation}
This has been applied in a self-gravitating disc to show that for larger sizes in an early disc ($\mathrm{St}=0.1$--$10$), gravitational collapse is suppressed by the internal diffusion of the dust particles \citep{Baehr2022}. For the size  $\mathrm{St} = 0.01$, we use values for the internal dust diffusion derived from the simulations of \citet{Baehr2021a} and shown in Table 1. In Figure \ref{fig:radialdustprofiles}, we plot $Q_{\mathrm{d}}$ as a function of radius for the three particle sizes for which we have measured dust diffusion coefficients.

The two larger dust sizes are below the critical stability threshold for large ranges of the radial extent of the disc. In particular, $\mathrm{St}=0.1$ grains are at or below the critical threshold $Q_{\mathrm{d}} = 1$, even for fairly stable discs. However, only beyond $\sim$ 100 au does the size $\mathrm{St}=0.1$ equate to millimetre grains, meaning that before accounting for some later planetary migration, grain growth above millimetre sizes may be necessary for the gravitational collapse of dust within 100 au.

The smallest size $\mathrm{St} = 0.01$ is above the critical value but represents an interesting case for slight variations in metallicity and concentration. In Figure \ref{fig:radialsmalldustprofiles}, we plot the curves for $Q_{\mathrm{d}}$ assuming that metallicity $Z$ is increased by a few above the assumed $Z = 0.01$. Even for slight increases to the metallicity, the curve for this small dust size can dip below the critical value. This does not take into account that the concentration factor $\epsilon$ is potentially larger due to increased metallicity and could cause the value of $Q_{\mathrm{d}}$ to decrease even further. Similarly, for $\mathrm{St}=1$, while the metallicity of this size is likely low due to the inefficiency of grain growth to this size, it can be potentially be overcome with the high concentration factors for dust of this size.

\begin{figure}
\centering
\includegraphics[width=0.48\textwidth]{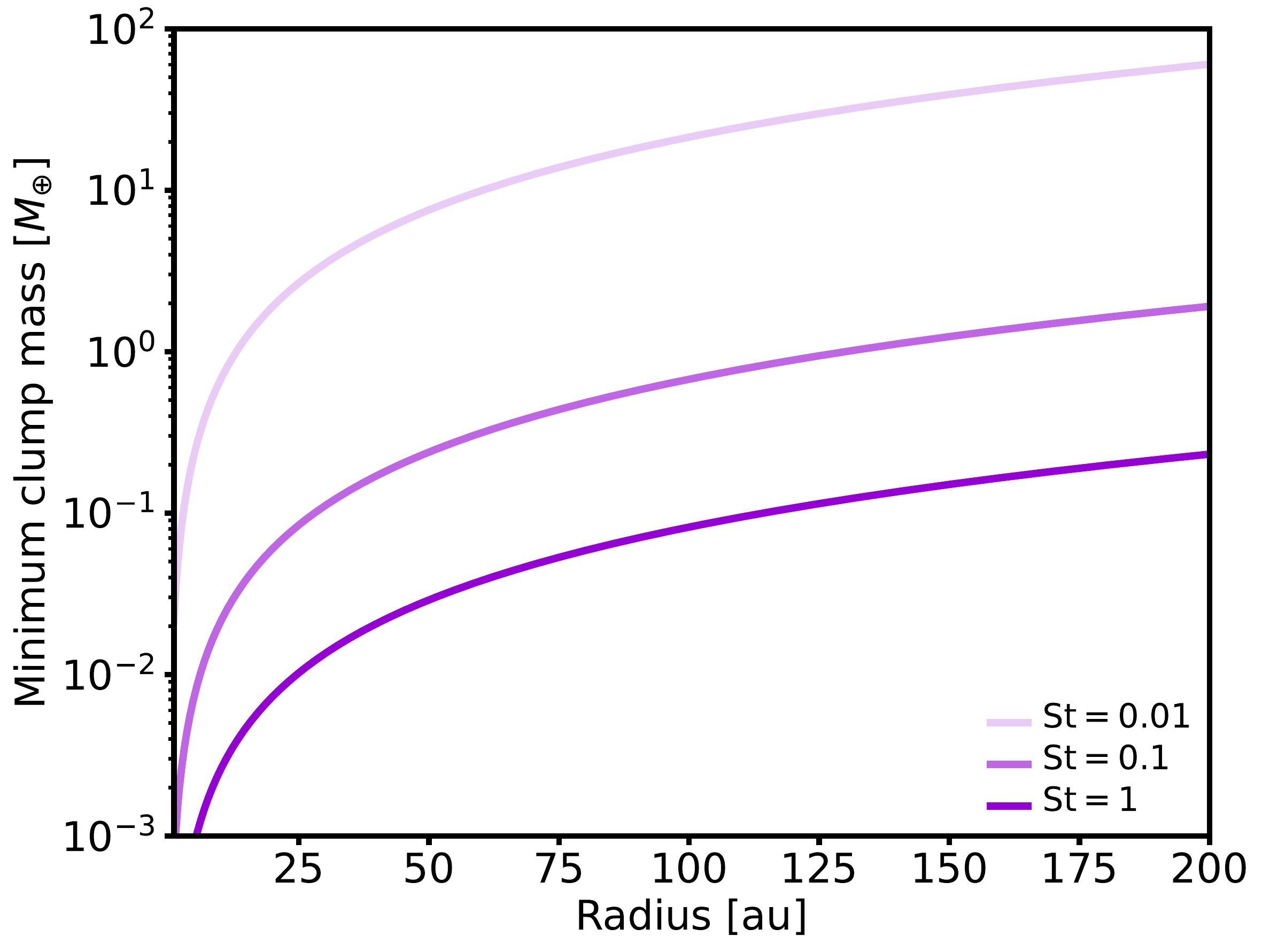}
\caption{The minimum mass necessary for the collapse of a cloud of dust comprised of a single particle size as derived from Equation \eqref{eq:characteristicmass}. Larger dust grains concentrate more readily and the mass required to overcome internal diffusion and reach Hill density is less than that of smaller grains.}
\label{fig:radialminimumclumpmass}
\end{figure}

Assuming enough dust exists that is susceptible to collapse, the gravitational collapse into bound clouds of dust depends on the internal diffusion of the dust grains. This process has a collapse timescale of 
\begin{equation} \label{eq:collapsetimescale}
\tau_{\mathrm{coll}} = \frac{1}{9\,\mathrm{St}}\Omega^{-1},
\end{equation}
for small dust $\mathrm{St} \leq 0.1$ and equal to the free-fall timescale for larger dust sizes at Hill densities, where self-gravity of the dust results in collapse into a bound cloud \citep{Klahr2020}
\begin{equation} \label{eq:freefalltimescale}
\tau_{\mathrm{ff}} = \sqrt{\frac{3\pi}{32G\rho}} \sim 0.64 \Omega^{-1}.
\end{equation}
Thus the formation of the dust clouds by gravitational collapse is among the shortest of the processes involved and will not be a limiting factor compared to the growth timescales.

The final mass of a gravitationally collapsing cloud of dust is inferred from the simulations of \citet{Baehr2022} which follow the non-linear evolution of dust concentration and collapse in a gravitoturbulent medium. Those results show that the the minimum mass regardless of particle species is around $10^{-2}\,M_{\oplus}$. Larger bodies up to a few $0.1 M_{\oplus}$ are possible for $\mathrm{St}=0.1$, and up to a few times $M_{\oplus}$ for the largest species $\mathrm{St}=1$. However, since these sizes are unlikely to be common in young discs, the focus will be on new simulations which include even smaller dust $\mathrm{St}=0.01$. Furthermore, there is very little overlap in the radial space where $\mathrm{St}=0.1$ dust particles are in the millimetre regime (see Figure \ref{fig:radialdustsize}) and the radial range where $\mathrm{St}=0.1$ dust is likely to collapse under its own self-gravity (Figure \ref{fig:radialdustprofiles}).

A cloud of dust must reach Hill density before it can gravitationally collapse. The critical radius $l_{c}$ of a cloud of dust depends on the strength of self-gravity and the internal diffusion \citep{Klahr2020,Klahr2021}
\begin{equation}
l_{c} = \frac{1}{3}\sqrt{\frac{\delta}{\mathrm{St}}} H_{g}.
\end{equation}
A cloud that is smaller than this critical radius will be unable to overcome the internal diffusion of its constituent particles and remain unbound. Larger than this radius and the self-gravity will be enough to collapse the cloud to one or more denser objects.

Thus for a given particle size $\mathrm{St}$, a cloud of particles at Hill density and of sufficient size $l_{c}$ will yield a minimum cloud mass that is bound
\begin{equation} \label{eq:characteristicmass}
M_{\mathrm{min}} = \frac{4\pi}{3} l^{3}_{c} \rho_{\mathrm{Hill}} = \frac{1}{9} \left( \frac{\delta}{\mathrm{St}} \right)^{3/2} \left( \frac{H_{g}}{R} \right)^{3} M_{\odot}.
\end{equation}
For clouds composed of the dust sizes of interest, we plot this mass as a function of radius in Figure \ref{fig:radialminimumclumpmass}. Since the Hill density is independent of particle size, the difference in minimum cloud size come from the turbulent properties of the dust. At our reference radius of 50 au this suggests that a seed formed with $\mathrm{St}=1$ sized dust would have a minimum mass of $4\times10^{-2}\, M_{\oplus}$ and $\mathrm{St}=0.1$ sized dust around $2\times10^{-1}\, M_{\oplus}$. The fact that smaller grains have a larger minimum mass is due to the fact that a cloud needs to be larger to overcome the dust diffusion at these sizes. For a disc with an outer cutoff of 200 au, gravitational collapse of dust into a bound object will likely results in objects within $R = 100$ au while a larger disc with a cutoff of 300 au can result in bound objects as far as 200 au.

From the simulations of \citet{Baehr2022}, gravitationally collapsed clouds of dust have merged and grown to these minimum masses and beyond such that clumps can range from $10^{-2}$ to $10$ Earth masses. However, the formation of bound clouds of dust was noticeably suppressed when using a smaller dominant species $\mathrm{St} = 0.1$, both in total number of bound clouds and total mass of the largest cloud, compared to the larger dust size $\mathrm{St}=1$.

In the following two sections, we consider that the seed for a planetary core has formed in the outer disc beyond $R = 20$ au. We assume this seed will be either be of $0.1 M_{\oplus}$ or $1 M_{\oplus}$ depending on how efficiently material from a $\mathrm{St}=0.1$ dust cloud can be converted to a solid object upon collapse. Simulations on the collapse of dust into planetesimals suggest over 80\% of material is immediately converted into number of solid bodies, although not necessarily into a primary object \citep{Polak2022}. Our assumption that solid seeds form depends heavily on the efficiency grain growth, which may vary greatly from one system to another and is by no means guaranteed. From there we consider the growth of the seed through pebble and gas accretion in a young, massive disc that is analogous to the three stages of growth for gas giant planet from \citet{Pollack1996}. First, the growth of pebbles will lead to a large enough core so that secondly, it can retain an gas atmosphere that will slowly grow as it cools through Kelvin-Helmholtz contraction. Once the core plus atmosphere has reached the crossover mass, the third stage, runaway gas accretion begins and the planet can quickly reach masses comparable to Neptune or Jupiter.

\section{Growth and Evolution in a Viscous disc}
\label{sec:evolution}

Starting with a solid object of $0.1$ or $1\, M_{\oplus}$, we can now begin the next stage in this process: accretion of solids up to a appreciable core. At this point the process is congruent to a core accretion model. Solid bodies such as those formed in \citet{Baehr2022} are in a gas rich environment with a modest gas atmosphere which will promote the accretion of more solids via pebble accretion \citep{Ormel2010}. Pebbles can be a more efficient means than planetesimals to grow up to the mass needed to grow up to the size of a ice or gas giant planet. Models of the formation of wide separation gas giants often require embryos of around a Martian mass at a few tens of au to reproduce the directly imaged planets at their current location \citep{Bitsch2015,Andama2022}.

\begin{figure*}
\centering
\includegraphics[width=0.48\textwidth]{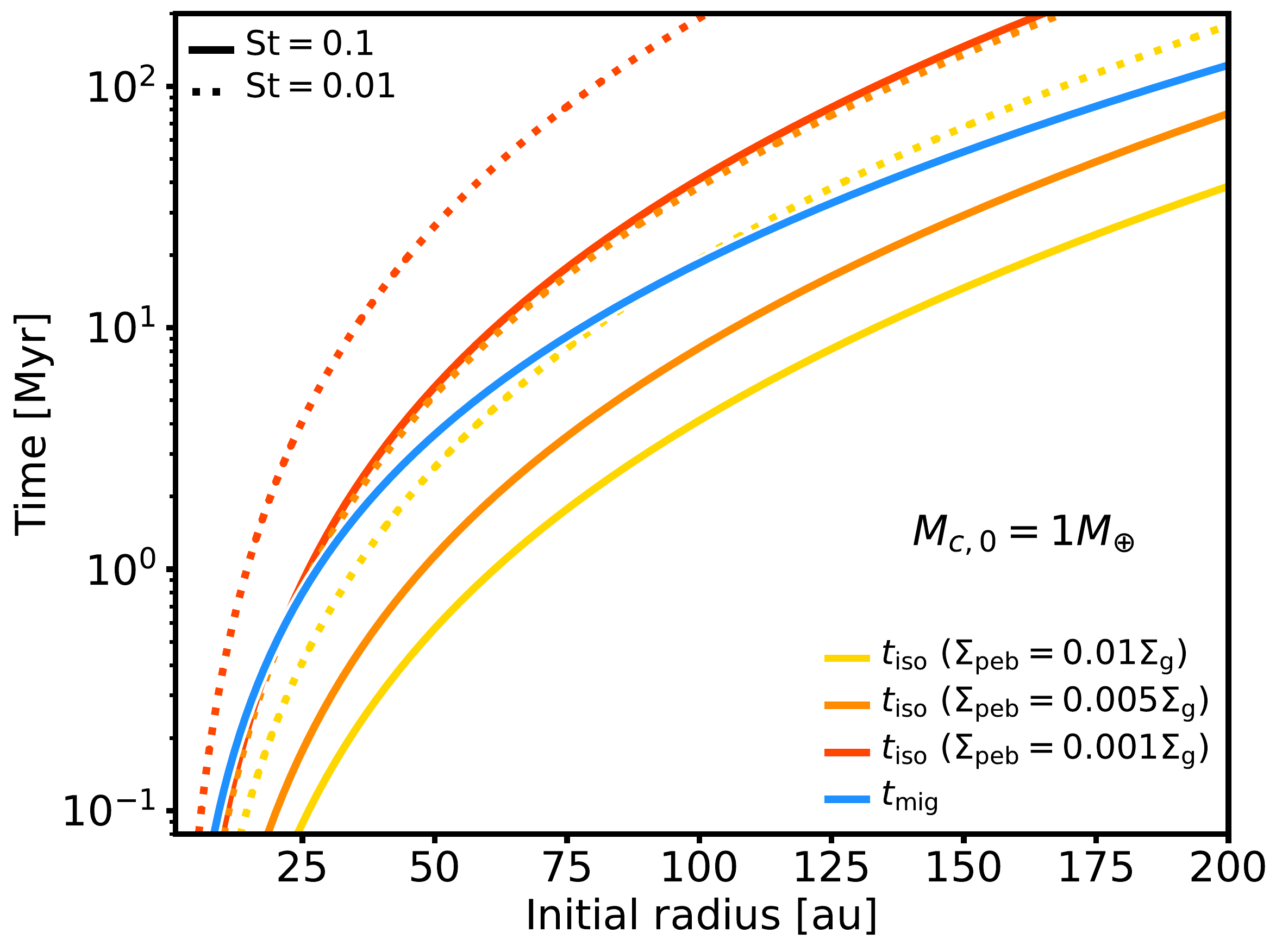}
\includegraphics[width=0.48\textwidth]{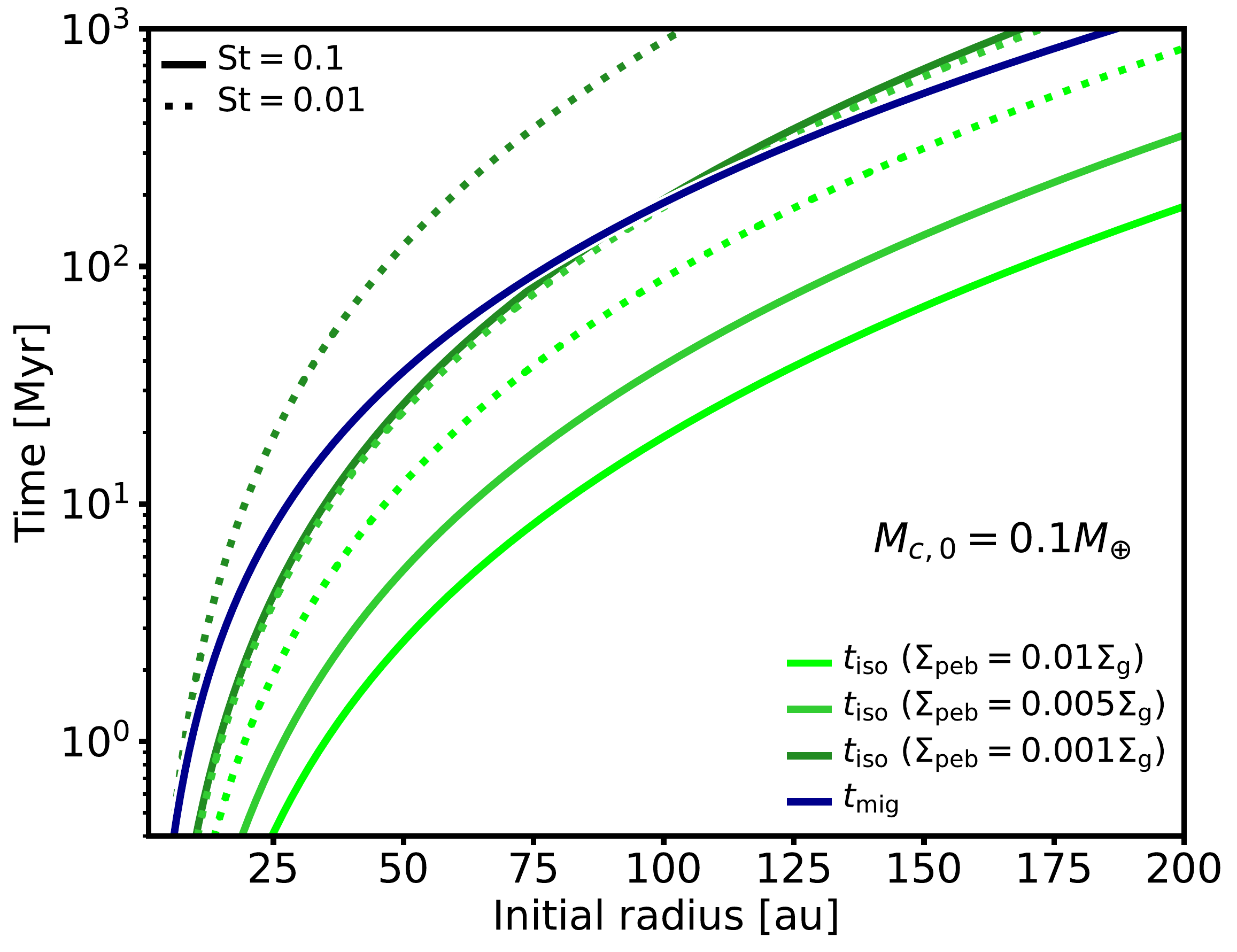}
\caption{If the pebble isolation mass is a critical limit in the growth of a planet, one can estimate a timescale to reach the isolation as a function of initial seed mass. We arrive at this timescale by dividing the pebble isolation mass of Equation \eqref{eq:pebbleisolationmass} by the accretion rate \eqref{eq:planet2Dpebbleaccretion} assuming an initial seed of $1 M_{\oplus}$ (left) and $0.1 M_{\oplus}$ (right). The three different solid curves assume the accretion of $\mathrm{St}=0.1$ pebbles with different for the pebble surface density as a function of the radial gas surface density. The dotted curves assume the accretion of $\mathrm{St}=0.01$ pebbles for the same pebble surface densities. In each case, the type-I migration timescale is overplotted (blue lines) as a potential constraint.}
\label{fig:radialpebbleisolationlimit}
\end{figure*}

\subsection{Accumulation of Solids}
\label{subsec:solidaccumulation}

Pebble accretion has been shown to accelerate the growth of planetesimals and embryos to planets \citep{Lambrechts2012}. Forming in the dense gas structures of a gravitationally unstable disc, we assume that our initial seeds have enough surrounding gas to decelerate passing pebbles. Pebble accretion generally falls into one of two regimes \citep{Johansen2017}, one where the accretion radius $R_{\mathrm{acc}}$ of a large body is smaller than the vertical height of the pebbles being accreted (3D regime)
\begin{equation}
\dot{M}_{\mathrm{3D,acc}} = \pi R_{\mathrm{acc}}^{2} \rho_{\mathrm{peb}} \delta v.
\end{equation}
and another where the accretion radius is larger than the pebble height (2D)
\begin{equation}
\dot{M}_{\mathrm{2D,acc}} = 2R_{\mathrm{acc}} \Sigma_{\mathrm{peb}} \delta v.
\end{equation}
The accretion cross-section of pebbles depends on the size of the accretor. For larger bodies, the $R_{\mathrm{Hill}}$ determines this cross-section, where
\begin{equation} \label{eq:hillradius}
R_{\mathrm{Hill}} = R \left( \frac{M_{c}}{3M_{*}} \right)^{1/3},
\end{equation}
is the Hill radius of a core with mass $M_{\mathrm{c}}$ at radius $R$. In this case, the approach velocity of the pebbles $\delta v = \Delta v + \Omega R_{\mathrm{acc}}$ is dominated by the shear velocity $\Delta v \ll \Omega R_{\mathrm{acc}}$. For objects greater than the transition mass $M_{t}$,
\begin{equation}
M_{t} = \sqrt{\frac{1}{3}} \frac{(\Delta v)^{3}}{G\Omega}
\end{equation}
the accretion radius is equal to the Hill radius, where the deviation from Keplerian velocity is
\begin{equation}
\Delta v = -\frac{1}{2} \frac{H_{g}}{R} \frac{\partial \ln p}{\partial \ln R} c_{s}.
\end{equation}
For the disc parameters assumed here, we find a transition mass equal to $2.5 \times 10^{-3}\, M_{\oplus}$ and the Hill regime is suitable for both seed masses.

If the disc remains self-gravitating, the scale height of small dust grains is able to settle a small amount due to the anisotropic transport of GI turbulence \citep{Riols2020,Baehr2021a}. However, for both accretor masses assumed here, 0.1 $M_{\oplus}$ and 1 $M_{\oplus}$, the scale height of these small pebbles is greater than the Hill radius. Thus, both seed masses are in the 3D regime and for  $R_{\mathrm{acc}} = (\mathrm{St}/0.1)^{1/3} R_{\mathrm{Hill}}$
%
\begin{align}
\dot{M}_{\mathrm{3D,acc}} &= \pi \left( \frac{\mathrm{St}}{0.1} \right)^{2/3} R_{\mathrm{Hill}}^2 \rho_{\mathrm{peb}} \Omega R_{\mathrm{Hill}}\\
&= \pi \left( \frac{\mathrm{St}}{0.1} \right)^{2/3} R^{3} \left( \frac{M_{c}}{3M_{*}} \right) \rho_{\mathrm{peb}} \Omega \label{eq:planet3Dpebbleaccretion}
\end{align}
%
In the this 3D regime, pebble accretion rates for the most optimistic assumption of the pebble fraction are less than $1\, M_{\oplus}/Myr$. This is insufficient to build up a sizeable core during the class 0/I stage, even if GI could be sustained for roughly million years and suggests that growth during the GI phase is limited. Therefore we limit growth through pebble accretion to the gravitationally stable disc, although there are additional considerations regarding evolution and migration in a gravitational unstable disc which we discuss in Section \ref{subsec:unstableevolution}. When the seed can accrete in the 2D limit, pebble accretion is faster by an order of magnitude, and thus the focus of the analysis.

If the disc is no longer self-gravitating, the height of the dust layer is smaller than the Hill radius and the mass accretion of pebbles is described by
\begin{align}
\dot{M}_{\mathrm{2D,acc}} &= 2 \left( \frac{\mathrm{St}}{0.1} \right)^{2/3} R_{\mathrm{Hill}} \Sigma_{\mathrm{peb}} \Omega R_{\mathrm{Hill}}\\
&= 2 \left( \frac{\mathrm{St}}{0.1} \right)^{2/3} R^{2} \left( \frac{M_{c}}{3M_{*}} \right)^{2/3} \Sigma_{\mathrm{peb}} \Omega \label{eq:planet2Dpebbleaccretion}
\end{align}
where $\Sigma_{\mathrm{peb}}$ is the pebble surface density. The pebble surface density can most easily be defined as the fraction of dust which is in pebbles $f_{\mathrm{peb}}$. Using the typical assumption for dust-to-gas ratios in a disc, we arrive at an expression similar to that of \citet{Forgan2019}
\begin{equation}
\Sigma_{\mathrm{peb}} \approx f_{\mathrm{peb}} Z\Sigma_{\mathrm{g}}
\end{equation}
The value of $f_{\mathrm{peb}}$ is uncertain and assumptions have ranged from $f_{\mathrm{peb}} = 0.1$ \citep{Forgan2019} to $f_{\mathrm{peb}}=0.5$ \citep{Ormel2017}, depending on the efficiency of grain growth.

For various rates of the amount of drifting pebbles with time, \citet{Bitsch2019} found that low pebble fluxes stunted gas giant planet formation while high pebble fluxes were able to rapidly grow the solid content of young embryos/cores. Young, self-gravitating discs likely still have a large reservoir of solids, meaning that higher pebble surface densities which favour rapid solid accretion are reasonable. While we keep a fixed pebble surface density, this will only be maintained by a equal or greater inward pebble flux. Pebble accretion rates for accreting bodies below $\sim 10\,M_{\oplus}$ are generally small enough that typical pebble fluxes are sufficient to maintain a steady pebble surface density profile \citep{Bitsch2019}.

When an accreting core is on an eccentric orbit, the relative velocity between core and pebbles can be significantly greater at the perihelion and aphelion ($\theta = \pi / 2$ and $3\pi /2$ respectively) than the Keplerian velocity \citep{Johansen2017}
\begin{align} \label{eq:eccentricity}
\Delta v &= ((ev_{K}\sin\theta)^2 + [(-1/2)ev_{K}\cos\theta + \eta v_{K}]^2 \nonumber \\
&+ (iv_{K}\cos\theta)^{2})^{1/2},
\end{align}
where $e$ and $i$ are the eccentricity and inclination, respectively, of the accreting body, $v_{K}$ is the disc Keplerian velocity and the pressure gradient term is defined as $\eta \equiv -(1/2) (H_{g}/R)^{2} (\partial \ln p /\partial \ln R)$ and is on the order of 0.01. For eccentricities around $e \sim 0.05$, this approaches the significance of the shear term $\Omega R_{acc}$ when the accreting body is at perihelion or aphelion and affect the accretion regime. Solid bodies like planetesimals can be excited up to eccentricities around $e=0.2$ \citep{Walmswell2013}, which means that at the perihelion and aphelion, the relative velocity can be 20\% and 10\% of the Keplerian velocity, respectively, and $\Delta v > \Omega R_{\mathrm{acc}}$. Thus for even moderate eccentricities and inclinations, accretion can instead occur within the smaller Bondi radius $R_{B} = GM/(\Delta v)^2$ near the apsides, and accretion rates will be reduced by a few orders of magnitude for either 2D or 3D accretion. For larger orbital radii, the damping timescale of eccentricity (or circularisation timescale) for $e=0.2$ can potentially become prohibitive for an Earth mass planet and disc mass $M_{D}$ \citep{Papaloizou2000}
\begin{align} \label{eq:eccentricitydamping}
t_{e} &\approx 1.3 \times 10^{5} \left[1 + \frac{1}{4}\left( \frac{e}{H_{g}/R} \right)^{3} \right] \left( \frac{H_{g}/R}{0.07} \right)^{4} \left( \frac{M_{D}}{2M_{\mathrm{Jup}}}\right)^{-1} \nonumber\\ 
&\times \left( \frac{M_{c}}{M_{\oplus}} \right)^{-1} \left( \frac{R}{50 au} \right)\, yr,
\end{align}
exceeding $10^{5}$ years beyond 130 au, which could delay the growth to the crossover mass. For the small particles that we accrete in this analysis ($\mathrm{St}<1$), eccentricities will be minuscule \citep{Shi2016} and unlikely to affect the accretion rate.

For lower planet masses ($\lesssim 0.3$ $M_{\mathrm{Jup}}$), inward radial migration is determined by the balance of Lindblad torques from the spiral waves of the embedded planet. The timescale for this so-called type-I migration can be parameterised via \citep{Tanaka2002}
\begin{align} \label{eq:migrationtime}
\tau_{\mathrm{mig,I}} &= C\frac{M_{*}}{M_{\mathrm{c}}}\frac{M_{*}}{\Sigma_{g} R^{2}} \left( \frac{H_{g}}{R} \right)^{2} \Omega^{-1}\\
&=2.79\times 10^{6} \left( \frac{M_{\mathrm{c}}}{M_{\oplus}} \right) ^{-1} \left( \frac{R}{50 au} \right) ^{2} \exp \left( -\frac{R}{R_{\mathrm{cut}}} \right) yr,
\end{align}
where the gas surface density $\Sigma_{g}$, and disc aspect ratio are at the radial location $R$ of a planet with mass $M_{\mathrm{pl}}$ and we assume a solar mass for the mass of the central star $M_{*} = M_{\odot}$. The constant $C$ accounts for the radial density and temperature structure of the disc, $C = 1/(2.5 + 1.7\beta - 0.1\alpha)$, where $\beta$ and $\alpha$ are the power law exponent of the radial temperature and density profiles, respectively. For the disc profiles we consider in Section \ref{sec:gravitoturbulentdiscs}, this value is $C=0.31$. Starting with the smallest initial seed size of $10^{-1} M_{\oplus}$, this timescale would be very long, longer than the lifetime of the gas disc for most initial radial locations. As the initial mass of the seed increases towards an Earth mass, the timescale becomes short enough that only the seeds that originate within a few tens of au might be expected to migrate onto the star.

Perhaps the first observable evidence for the growth of a planet in a disc is the cessation of dust inward drift outside of the orbit of the planet, which can create dust gaps in a circumstellar disc. This occurs once the companion has reached the pebble isolation mass $M_{\mathrm{iso}}$ \citep{Lambrechts2014}, which can be defined in terms of the gas pressure scale height \citep{Bitsch2018} such that an isolation mass of 
\begin{equation} \label{eq:pebbleisolationmass}
M_{\mathrm{iso}} \approx 20\left( \frac{H_{g}/R}{0.05} \right)^{3} M_{\oplus},
\end{equation}
halts the inward drift of pebbles and limits the amount of solids that can be deposited on the embedded planet. Using the solid accretion rate in Equation \eqref{eq:planet2Dpebbleaccretion}, we derive a time to reach the pebble isolation mass $t_{\mathrm{iso}} = M_{\mathrm{iso}}/\dot{M}_{\mathrm{2D,acc}}$ and compare it against the type I migration rate in Figure \ref{fig:radialpebbleisolationlimit}. We plot the timescales to reach the pebble isolation mass with different values of $f_{\mathrm{peb}}=0.1,0.5,1$, corresponding to a range of pebble surface densities $\Sigma_{\mathrm{peb}}$ and assuming accreted pebbles are either $\mathrm{St}=1$ or $\mathrm{St}=0.1$. This is a minimum mass estimate because the pebble accretion rate at all radii is approximated by the accretion rate at the origin, which will increase as a core migrates radially inwards. Furthermore, this assumes that the seed migrates through the entirety of the disc, but once the pebble isolation mass is reached, the supply of pebbles will ebb and limit further growth.

\begin{figure*}
\centering
\includegraphics[width=0.48\textwidth]{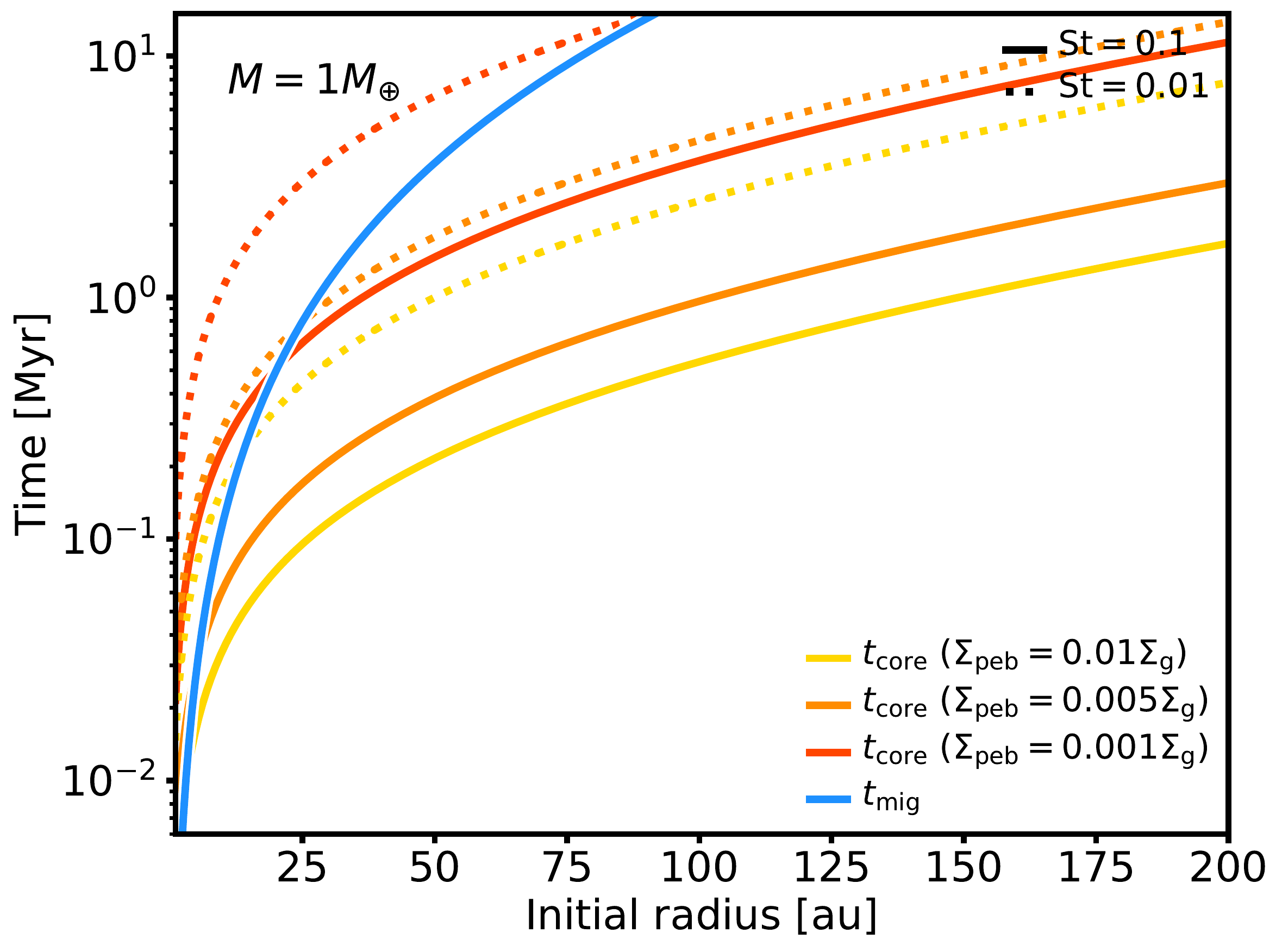}
\includegraphics[width=0.48\textwidth]{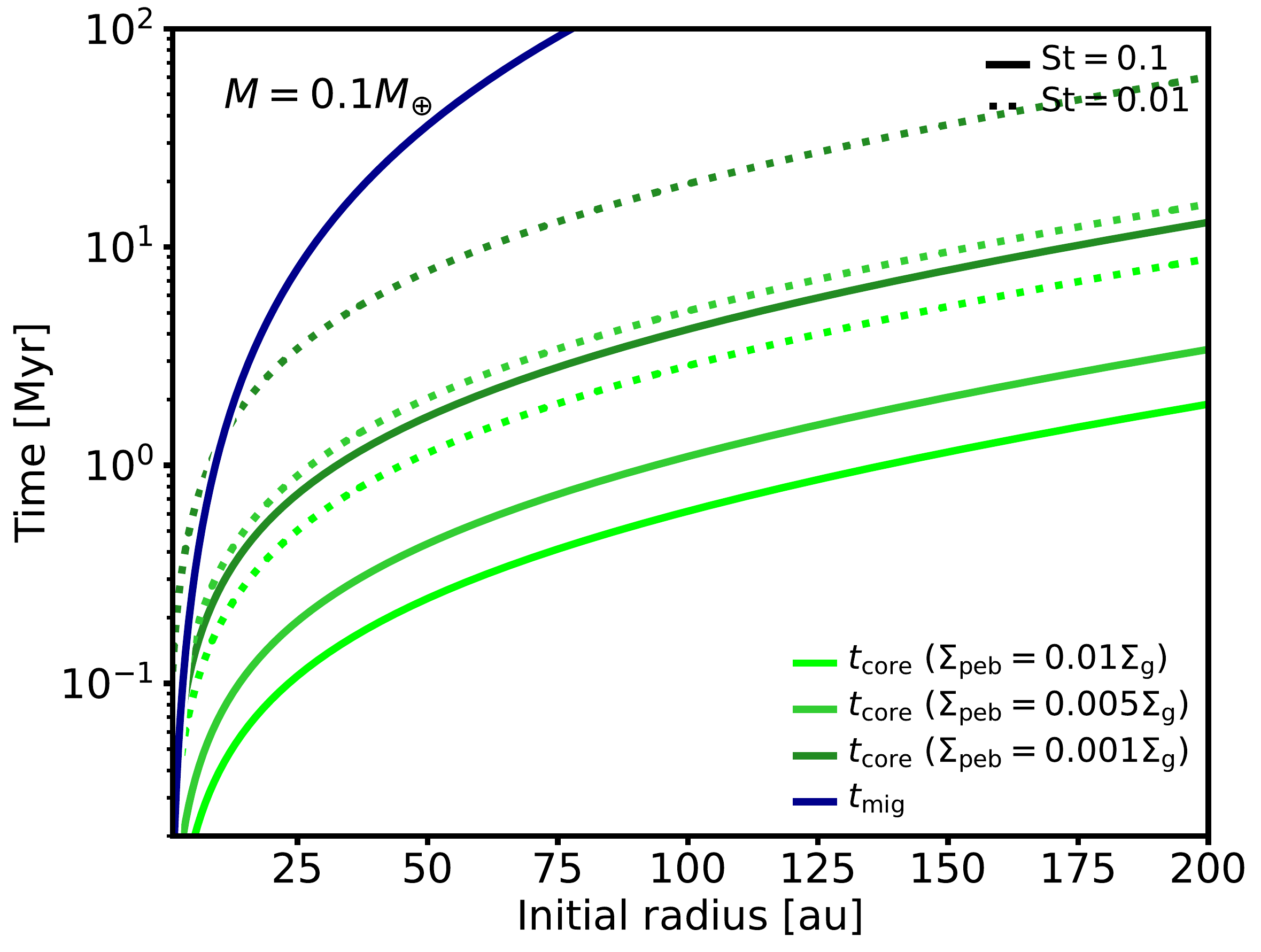}
\caption{Same as Figure \ref{fig:radialpebbleisolationlimit}, but using the runaway mass from Equation \eqref{eq:crossovermass} as the target threshold at which point the presence of a growing planet in a disc can be inferred. Reaching the crossover mass is much easier in the outer disc than reaching the pebble isolation mass, making the timescales for gas giant formation much more feasible.}
\label{fig:radialrunawaylimit}
\end{figure*}

While the timescales to reach the pebble isolation mass in Figure \ref{fig:radialpebbleisolationlimit} are short enough to grow to the pebble isolation mass before migrating on to the star, they also approach the lifetime of the typical gas disc. Thus we require a different approach to reach a planet mass milestone that can be inferred through observations.

Alternatively, one can ask what is the minimum core mass required to reach runaway growth, as was done in \citet{Ormel2021}. This assumes that pebble accretion, and the associated luminosity, sets the crossover mass threshold $M_{\mathrm{cross}}$, and that accreted pebbles are not necessarily deposited on the core but possibly evaporated into the atmosphere \citep{Brouwers2020}. The crossover mass then depends on the planet radius where silicates transition from solid to vapour, parameterised by the vapour temperature of the accreted silicate pebbles $T_{\mathrm{vap}}$, the opacity at the radiative-convective boundary $\kappa_{\mathrm{rcb}}$, in addition to the orbital radius, seed mass and pebble accretion rate.
\begin{align} \label{eq:crossovermass}
M_{\mathrm{cross}} &= 10 \left( \frac{\kappa_{\mathrm{rcb}}}{0.01 g cm^{-2}} \right)^{1/6} \left( \frac{R}{50\,au} \right)^{7/108} T_{\mathrm{disc}} ^{-7/27} \nonumber \\
&\times\left( \frac{T_{\mathrm{vap}}}{2500 K} \right)^{16/27} \left( \frac{\dot{M}_{\mathrm{2D,acc}}}{10^{-5} M_{\oplus} yr^{-1}} \right)^{1/6} \left( \frac{M_{c}}{M_{\oplus}} \right)^{1/2} M_{\oplus}
\end{align}
We simplify the above equation by assuming that the opacity and vapour temperature are a constant $\kappa_{\mathrm{rcb}} = 0.01$ and $T_{\mathrm{vap}} = 2500$, respectively.
\begin{equation} \label{eq:crossovermass2}
M_{\mathrm{cross}} = 5 \left( \frac{R}{50\,au} \right)^{7/36} \left( \frac{\dot{M}_{\mathrm{2D,acc}}}{10^{-5} M_{\oplus}/yr} \right)^{1/6} \left( \frac{M_{c}}{M_{\oplus}} \right)^{1/2} M_{\oplus}
\end{equation}
The timescales $t_{\mathrm{core}} = M_{\mathrm{cross}}/\dot{M}_{\mathrm{2D,acc}}$ for this new threshold mass are plotted in Figure \ref{fig:radialrunawaylimit}. The time to reach the crossover mass is still shorter than the migration timescale in the outer disc, but now it is significantly faster to reach the runaway mass than the pebble isolation mass. This is particularly true at large radii, where the crossover mass can be reached within a million years even for the more conservative assumptions for pebble abundances. The accretion rate is based on the initial seed mass and thus as it grows the accretion rate will increase and the timescale to reach the crossover mass becomes shorter. Therefore the timescales presented here likely overestimate $t_{\mathrm{core}}$, but it can also be affected by changes in disc location, migration rate, gas and pebble densities, etc. Understanding how all these factors effect eventual planet size and location is left to future numerical experiments.

\subsection{Evolution towards a Gas Giant}
\label{subsec:gasgiantgrowth}

After the core plus atmosphere has reached the crossover mass, continued growth is roughly determined by the amount of gas that can be gravitationally bound to the embedded planet. We now consider the core plus atmosphere to be a planet and so we replace $M_{\mathrm{c}}$ with $M_{\mathrm{pl}}$. The mass accretion rate will again depend on two critical length scales, the Hill radius \eqref{eq:hillradius} and the Bondi radius \citep{DAngelo2008}. The Bondi radius $R_{\mathrm{B}}$ is the limit at which the gas velocity is equal to the escape velocity of the planet
\begin{equation}
c_{s}^{2} = \frac{2GM_{c}}{r},
\end{equation} 
where $r$ is radial distance from the planet centre. The yields the critical radius
\begin{equation} \label{eq:bondiradius}
R_{B} \equiv \frac{2GM_{\mathrm{pl}}}{c_{s}^{2}}.
\end{equation}

For an embedded planet in a disc, a simplified model for gas accretion means that all gas within the Bondi radius is captured by the planet. This is unlikely to be the full picture, as 3D hydrodynamic simulations suggest there is significant mixing and recycling of material between the planet and disc \citep{Cimerman2017}. Simulations of atmospheric recycling generally assume conditions (i.e. temperature, opacity) typical of the inner disc, which may not directly translate to outer disc conditions. Nevertheless, for the purposes of a first order estimate, the accretion rate can be approximated by the gas density that passes through the cross sectional area $\pi r^2$ with a relative velocity $\Omega r$ with gas density $\rho_{g} = \Sigma_{g}/H_{g}$ \citep{DAngelo2008}
\begin{equation}
\dot{M} \sim \frac{\Sigma_{g}}{H_{g}}\pi \Omega r^{3}.
\end{equation}
For smaller planet masses where $R_{B} < R_{H}$, this amounts to $r = R_{B}$. 
\begin{align}\label{eq:runawaytimescale}
\dot{M}_{B} &= \frac{\Sigma_{g}}{H_{g}}\pi \Omega \left( \frac{GM_{\mathrm{pl}}}{c_{s}^{2}} \right)^{3}\\
&= 0.012 \left( \frac{R}{50\, au} \right)^{-11/4}  \left( \frac{M_{\mathrm{pl}}}{10 M_{\oplus}} \right)^{3} \exp \left( -\frac{R}{R_{\mathrm{cut}}}  \right) M_{\oplus}\, yr^{-1}
\end{align}
This accretion rate assumes that the disc has the high gas surface densities near gravitational instability, which could lead to very high accretion rates.

The point at which the Bondi radius and the Hill radius are equal is commonly called the disc thermal mass and is also the point Hill radius becomes comparable to the disc scale height $H_{g}$ \citep{Goodman2001,Rafikov2006}
\begin{equation} \label{eq:thermalmass}
M_{\mathrm{th}} \approx \frac{c_{s}^{3}}{\Omega G} = 33 \left( \frac{R}{50\, au} \right) ^{3/4} M_{\oplus}.
\end{equation} 
This is the point where the planet may begin to open up a gap, separate from the disc and spirals become non-linear shocks \citep{Zhu2015,Bae2018a}. Inside a few au the thermal mass and the crossover mass are comparable, but beyond around 10 au, this critical core mass for runaway growth is reached before the thermal mass. Thus by the time the thermal mass has been reached in the outer disc, runaway growth is well underway and planet masses can quickly reach gap opening limits.

\begin{figure}
\centering
\includegraphics[width=0.48\textwidth]{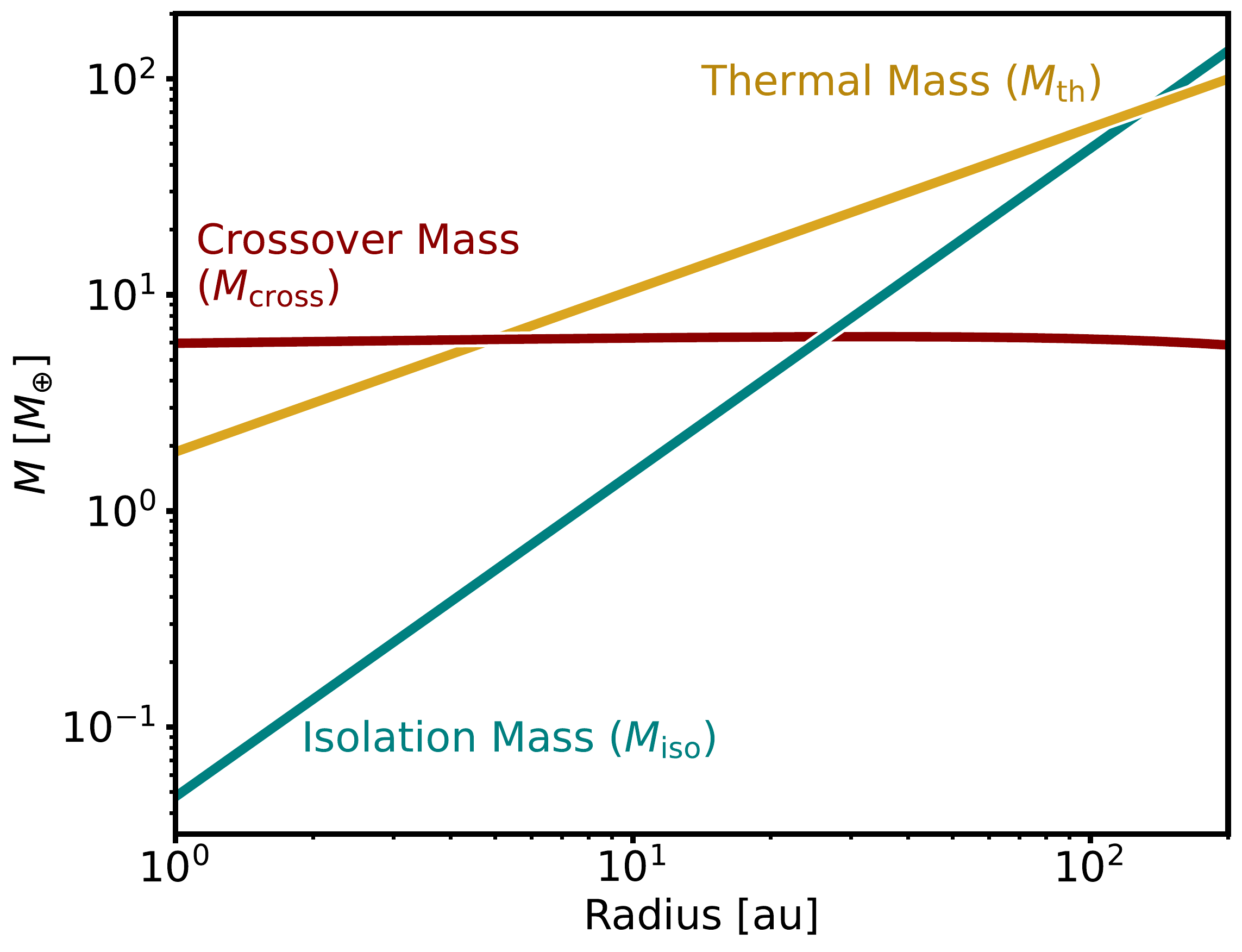}
\caption{Important planetary mass thresholds as a function of disc radius. In the outer disc the crossover mass necessary for the onset of runaway growth is smaller than the disc thermal mass and the pebble isolation mass.}
\label{fig:radialmassthresholds}
\end{figure}

\begin{figure*}
\centering
\includegraphics[width=0.75\textwidth]{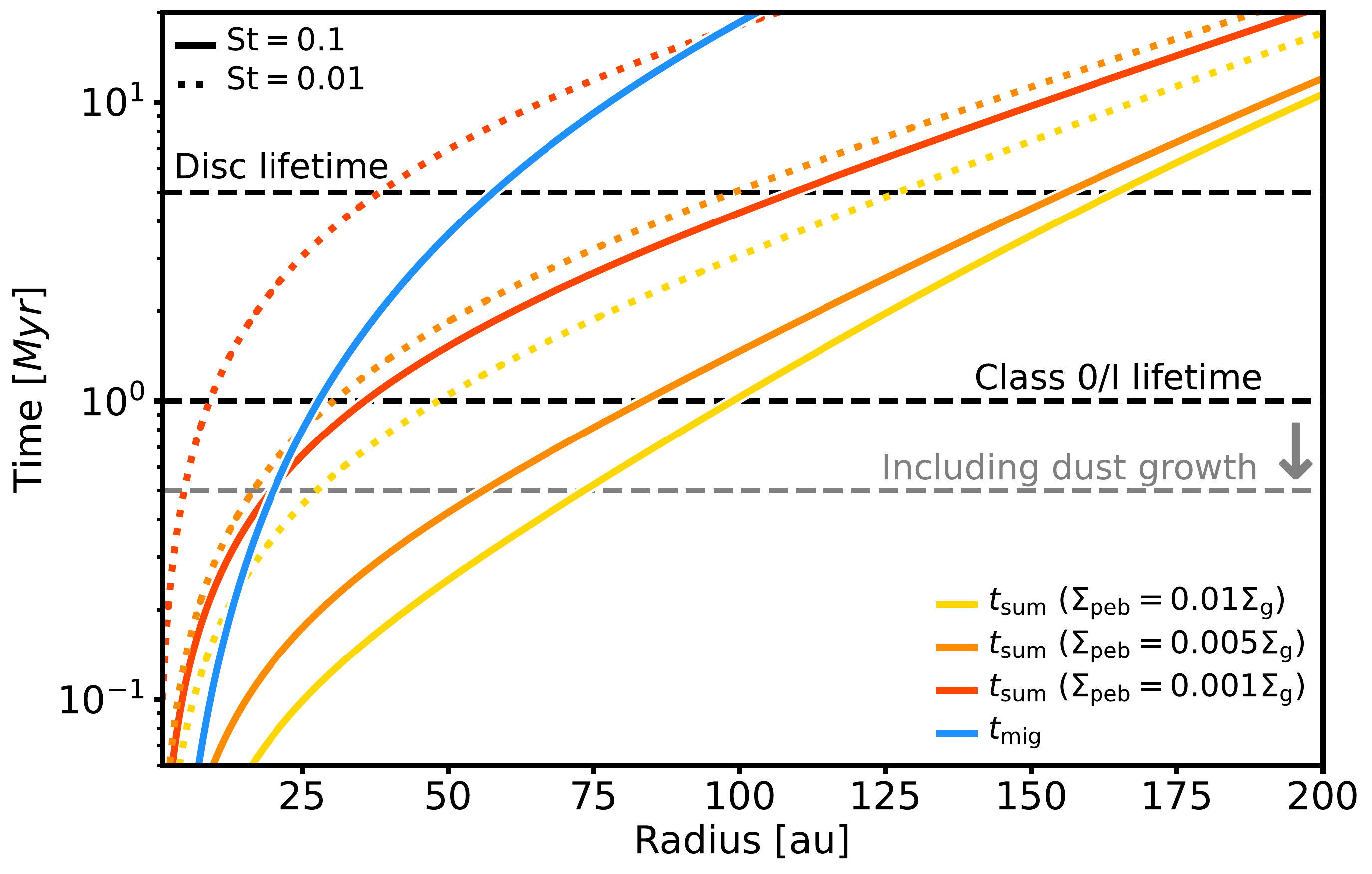}
\caption{Here we add together the two timescales used in this paper: (1) the time needed to reach a sizeable core (crossover mass) $t_{\mathrm{core}}$, and (2) time time to reach the thermal mass during the runaway phase $t_{\mathrm{th}}$, for different pebble surface densities as a function of radius. We use the model that assumes that we begin with a $1 M_{\oplus}$ seed that accretes pebbles of size $\mathrm{St} = 0.1$ (solid lines) or $\mathrm{St} = 0.01$ (dotted lines). Plotted as horizontal dashed lines are constraints on these timescales, including the lifetime of the gas disc $t \sim 5 \times 10^6$ years and the rough lifetime of a class 0/I disc $t \sim 1 \times 10^{6}$ years. We also include the a grey dashed line for a class 0/I disc assuming that half the time is required for dust growth, $t \sim 5 \times 10^{5}$ years. Solid lines indicate that pebbles accreted during stage (1) are size $\mathrm{St}=0.1$ and dashed lines are for the accretion of $\mathrm{St}=0.01$ pebbles.}
\label{fig:summation}
\end{figure*}

If we are interested in the time necessary to reach the thermal mass when $R_{B} < R_{H}$ we calculate $t_{\mathrm{th}} = M_{\mathrm{th}}/\dot{M}_{B}$ as
\begin{equation}
t_{\mathrm{th}} = 2750 \left( \frac{R}{50\, au} \right)^{7/2}  \left( \frac{M_{\mathrm{pl}}}{10 M_{\oplus}} \right)^{-3} \exp \left( \frac{R}{R_{\mathrm{cut}}}  \right) yr
\end{equation}

At high enough planet masses a gap can be cleared in the disc and migration will fall into the type-II regime, where the transfer of angular momentum of gas in horseshoe orbits within the co-rotation region leads to radial migration \citep{Kley2012}. This can generally be described by inward radial migration on a timescale of \citep{Lodato2019}
\begin{equation}
t_{\mathrm{mig,II}} = \frac{2}{3\alpha} \left( \frac{H_{g}}{R} \right)^{-2}\Omega^{-1}.
\end{equation}
Typical values of $\alpha$ in a gravitationally stable disc range from $10^{-3}$ to $10^{-5}$, which is supported by observations \citep{Flaherty2015,Flaherty2017}, such that type-II migration will operate on timescales of hundreds of thousand to millions of years, which is significantly longer than the runaway growth times. Therefore, type-II migration will have a smaller effect than type-I migration as a constraint on planet formation.

\begin{figure}
\centering
\includegraphics[width=0.48\textwidth]{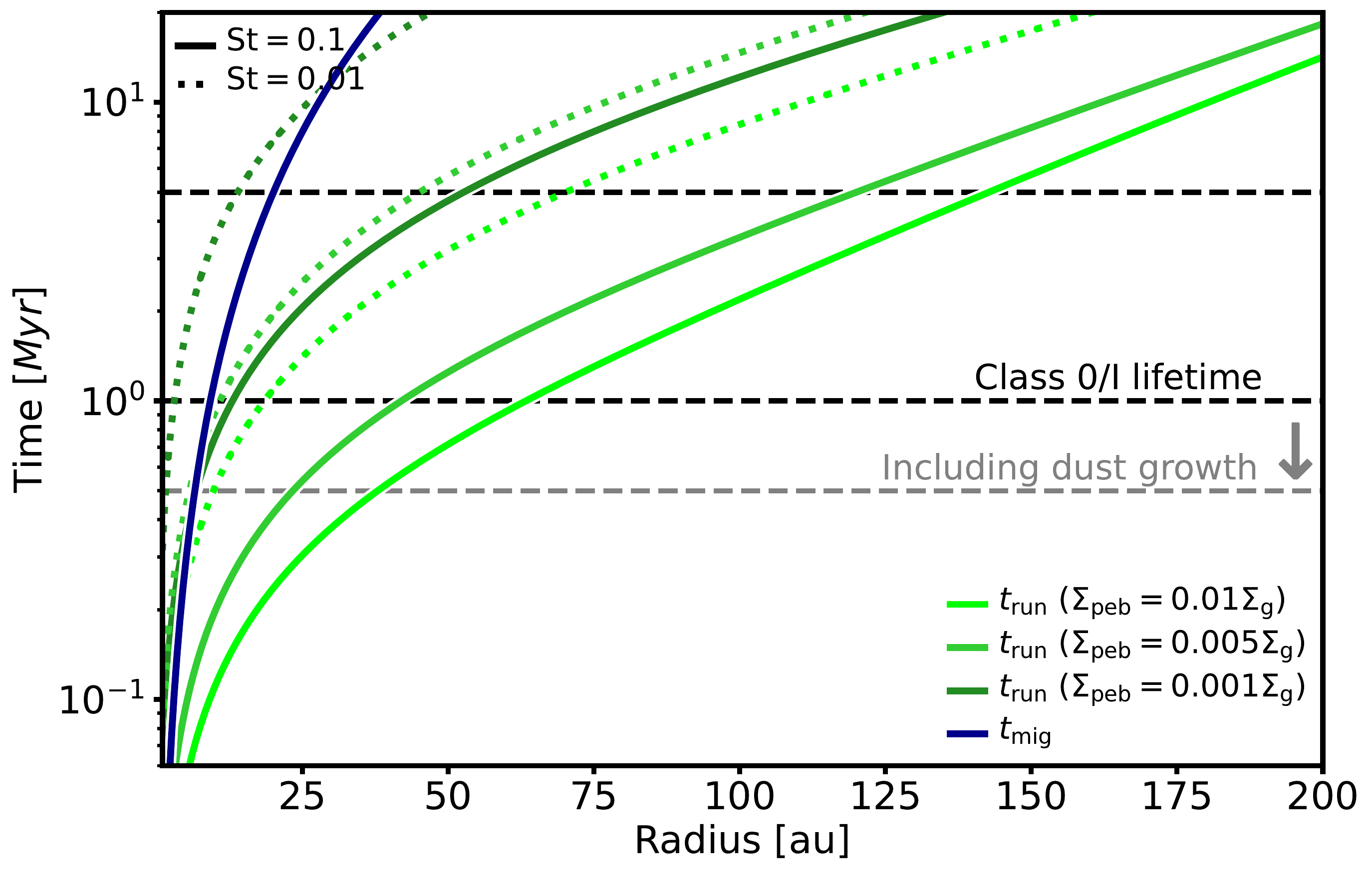}
\caption{The cumulative formation timescale as in Figure \ref{fig:summation} but for an initial seed mass of $0.1\, M_{\oplus}$. Pebble accretion in this case is less efficient and formation of a core with the crossover mass is only possible with approximately 50 au, for generous assumptions.}
\label{fig:summationsmallcore}
\end{figure}

We plot a cumulative formation timescale using the timescales $t_{\mathrm{sum}} = t_{\mathrm{core}} + t_{\mathrm{th}}$ associated with the formation of a gas giant in Figure \ref{fig:summation}: (1) the time needed to reach a sizeable core $t_{\mathrm{core}}$ and (2) the time time to reach the thermal mass during the runaway phase $t_{\mathrm{th}}$. The regions where the total formation timescale is less than $\sim 1$ Myr or roughly within the lifetime of a class 0/I disc depends largely on the efficient accretion of pebbles $t_{\mathrm{core}}$. In the far outer disc ($>$ 100 au) pebble accretion and runaway accretion both become less efficient and formation timescales become larger than the lifetime of the gas disc $t_{\mathrm{disc}} \sim 5 \times 10^{6}$ years. When only $\mathrm{St} = 0.01$ pebbles can be accreted on to a $0.1\, M_{\oplus}$ seed, the overall formation time is nearly an order of magnitude longer and barely possible within 1 Myr (see Figure \ref{fig:summationsmallcore}).


\subsection{Evolution in a Gravitationally Unstable Disc}
\label{subsec:unstableevolution}

So far we have considered the case where the disc is viscous but not considerably self-gravitating $Q > 2$. In the case that mass infall onto the disc is greater or equal to the mass lost to accretion onto the star and any mass surrounding the growing planet, the disc can remain unstable. If the disc remains unstable for much longer after the formation of a solid embryo, the fate of the body is much more uncertain. Migration rates of gas giants planets assumed to have formed by fragmentation have been shown to migrate on very short timescales \citep{Baruteau2011,Zhu2012}, however it may be possible to halt the migration \citep{Rowther2020} when cooling is not constant throughout the disc. Furthermore, while self-gravity can make it easier to open up a gap in a disc, for high surface densities approaching $Q=1$, turbulence can suppress the formation of gaps \citep{Zhang2014}.

However, what we consider here is not gas giants, but rocky embryos or cores with tenuous atmospheres, which is not often considered in the literature because GI has long been associated with direct formation of gas giants. \citet{Baruteau2008} considered how objects that are too small to open up a gap in the disc can move in a self-gravitating disc, i.e. type I migration, and found that disc self-gravity makes type I migration faster. However, type I migration is still faster than type II migration by almost an order of magnitude, as suggested in non-linear simulations of planets as low as 20 $M_{\oplus}$ \citep{Baruteau2011}. Objects smaller than one Earth mass are too small to launch spiral arms at Lindblad resonances and are potentially subject to repeated interactions with the spiral arms of the self-gravitating disc \citep{Boss2013}. Without a robust understanding of the long term growth and evolution of rocky embryos and cores in a gravitoturbulent disc, for now we assume that the formation of rocky objects coincides with the end of strong accretion.

\section{Discussion}
\label{sec:discussion}

One main advantage of this scenario over traditional planet formation via disc fragmentation is that it can greatly accelerate the planet formation process in the outer regions of the disc. It does so by bypassing a few important steps to forming an embryo in the traditional core accretion pathway.

For now, we can now begin to roughly put these parts together, the growth and concentration of dust, the accretion of solids and gas, into a planet formation pathway that can rapidly produce planets on wide orbits. To do this we need to limit the discussion to two idealised cases, one where dust can efficiently grow to millimetre sizes (roughly approximated by $\mathrm{St}=0.1$ in the outer disc) and the other where dust is limited to largely dust grains of around 100 microns in size ($\mathrm{St}=0.01$). While larger dust on the order of centimetres offers the most efficient trapping properties, growth to these sizes in the outer disc are unlikely due to the efficiency of radial drift (Figure \ref{fig:radialdustsize}).

For the case that dust does not efficiently grow, there are two obstacles to rapid planet formation. First, the total mass required to overcome the internal diffusion of the dust and form a bound object is higher (see Figure \ref{fig:radialminimumclumpmass}), to the point where one needs an entire super-Earth's worth of solid material. Rapid though this may be, concentration of this magnitude for dust so well-coupled to the gas is unlikely. Second, even in the most optimistic assumption of pebble surface densities, the accretion of dust of this size up to the isolation mass is on the order of the lifetime of a typical protoplanetary disc. To avoid forming a planetary core in a disc devoid of pebbles to accrete one must assume dust growth to larger sizes occurs after the formation of the solid seed.

Assuming that dust can indeed grow to millimetre sizes, dust concentrates within dense gas structures more easily resulting in initial seed masses are closer to the mass of the Earth. We plot the sum of the timescales necessary to reach the necessary size limits in Figure \ref{fig:summation}. This is already a mass two orders of magnitude larger than the seed embryos employed by pebble accretion models \citep{Bitsch2019,Andama2022}. Crucially, unlike the case where dust does not grow up to millimetre sizes, the accretion of millimetre dust grains is much more efficient and makes the whole formation process feasible on timescales of less than $1\times 10^{6}$ years within 100 au.

\begin{figure}
\centering
\includegraphics[width=0.48\textwidth]{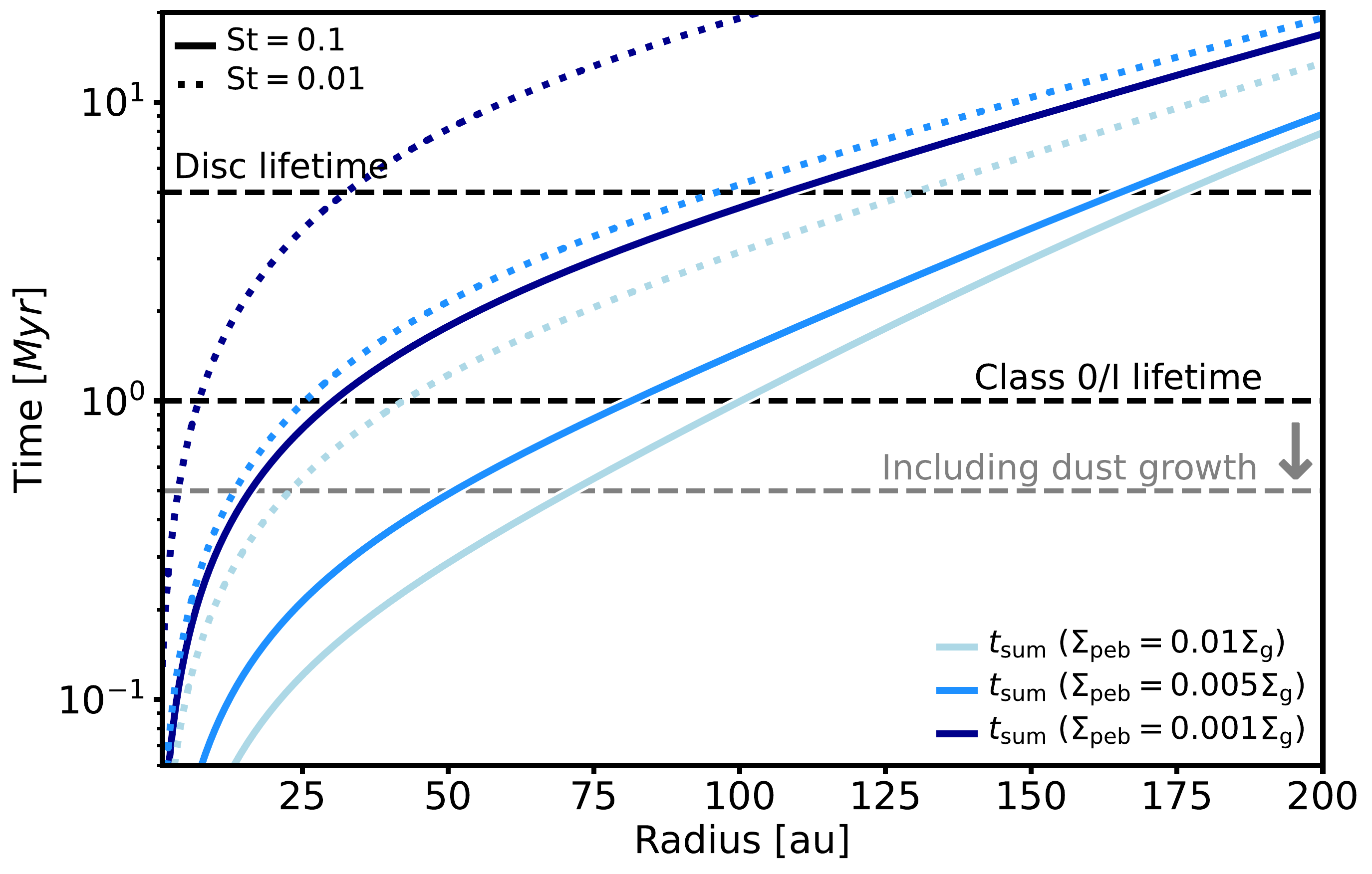}
\caption{The cumulative formation timescale as in Figure \ref{fig:summation} but for an outer disc cutoff radius of 300 au instead of the previously assumed 200 au. This choice does not have significant impact on the cumulative formation timescale. Inside 100 au, gas surface densities are only modestly where efficient pebble accretion can rapidly build a core, leaving the cumulative formation time largely unchanged. Outside of 100 au, the gas densities are higher, but not enough to significantly reduce the cumulative formation timescale.}
\label{fig:summationlargedisk}
\end{figure}

The curves in Figure \ref{fig:summation} assume that the disc has an exponential cutoff radius of 200 au, but the summed timescales are almost identical if one assumes a cutoff radius of 300 au. Inside of 100 au, where the efficient accretion of pebbles keeps the overall formation timescale reasonably short, the marginally higher gas surface density does not result in significantly higher pebble accretion rates. Outside of 100 au, the higher gas density decreases the time to reach the thermal mass, but only by a factor around unity, which is not enough to dramatically alter the range where gas giants may form (see Figure \ref{fig:summationlargedisk}).

\subsection{Reconciling with Current Theories of Planet Formation}
\label{subsec:alternateformation}

The formation of metre-sized objects has long been a obstacle to planet formation due to fragmentation and radial drift barriers \citep{Birnstiel2012}, both of which are overcome here by forming the embryo or core directly from small dust grains. The manner of subsequent growth of the planet through the accretion of gas and dust is undiminished by forming in the first million years when the gas disc is still intact. It has been suggested that all companions with masses less than a few Jupiter masses belong to a population which have metallicities more consistent with core accretion \citep{Schlaufman2018}. The scenario proposed here would remain consistent with a picture that disc fragmentation create brown dwarfs and low mass stars and Jupiter mass objects are formed from an initial solid core. This is potentially supported by the separation distribution of binary stars and gas giants, which are very different especially with respect to stellar metallicity, suggesting that stellar companions form differently from gas giant planets \citep{Offner2022}.

We use the formation of planetary embryos in a gravitoturbulent disc principally as a means to form the distant ($R > 30$ au) planet that grow up to masses between 1 and 10 $M_{\mathrm{Jup}}$ within a few million years. This observational space is where direct imaging has detected a number of gas giant planets and a number of discs have been observed with gaps at large radii \citep{Andrews2018a,Long2018,Clarke2020,Segura-Cox2020,Sheehan2020} that might host planets \citep{Pinte2020}. If these gaps are carved by growing planets, most are small enough to be outside of current detection limits of direct imaging \citep{Zhang2018,Lodato2019}.

While streaming stability is not especially efficient at large radial distances, \citet{Jang2022} suggest that rapid accretion of solids via pebble accretion could create a gap-opening planet within 1 to 3 million years. This however is only possible if the disc is assumed to be at or near the gravitational instability limit, which is unlikely for extended periods of time $>$ 1 million years. Starting with a planetary embryo at a very early stage could shorten the time needed to be near the gravitational instability.

The rapid formation of cores has been proposed to form planets within disc substructures, using streaming instability \citep{Cridland2022,Lau2022}. This scenario requires similar amounts of grain growth in a young disc, but less local concentration, since the formation of planetesimals is still an important intermediate step. It may also require the existence of axisymmetric pressure bumps to facilitate grain growth and concentration \citep{Carrera2021}.

Secular gravitational instability (SGI) has been proposed as a means to concentrate dust in axisymmetric rings and potentially form planetesimals at large radial separations \citep{Michikoshi2012,Takahashi2016}. This can produce embryo sized objects similar to those of GI \citep{Takahashi2022} with a similar dust stability criterion as Equation \eqref{eq:gerbigparameter} \citep{Latter2017,Tominaga2020}.

An alternative means of rapidly forming intermediate-sized planets returns to the idea of disc gravitational instabilities, but considers the effect of magnetic fields as a means to suppress the fragment size so that Neptune-sized planets are formed \citep{Deng2021}. Earlier work on the interplay between gravitational instabilities and magnetohydrodynamics (MHD) found that magnetic fields increase the pressure support against collapse and increase the mass of disc fragments \citep{Forgan2017}. That work perhaps lacked the resolution to capture the MHD turbulence and with improved resolution \citet{Deng2021a} found that magnetic fields shield the fragment from further accretion and growth.

While wide separation gas giants have been detected around older systems which no longer have a circumstellar disc, there are a growing number of gas giants candidates at wide separations and within the discs of very young systems which may require a particularly efficient mechanism to explain their formation. Gas kinematics of HD 163296 \citep{Pinte2018} and IM Lup \citep{Verrios2022} suggest that gas giants of a few $M_{\mathrm{Jup}}$ are embedded at radii around 100 au. Formation via traditional core accretion is difficult to reconcile at these disc ages and radial locations, but much more feasible considering the scenario we propose.

Some observations of young systems indicate that dust is potentially concentrating by some mechanism. The IRAS 16293-2422 A \citep{Maureira2020} and L1489 \citep{Ohashi2022a} systems show that there are potentially large dusty regions around very young central sources. Similarly, the structure of the dust emission in L1448 IRS3B \citep{Tobin2016} shows a large quantity of dust but with a clearer structure. The detection suggests a large quantity of dust is concentrating from a spiral feature, unusual since disc fragmentation is dominated by the collapsing gas rather than the dust, although the dust could be strongly affected by the collapsing gas \citep{Boley2010,Baehr2019}.

\subsection{Future Considerations}
\label{subsec:futureconsiderations}

A number of parameters that govern the gravitational collapse of dust in Equation \eqref{eq:gerbigparameter} are only loosely constrained in young discs. While the concentration factor $\epsilon$ is measured from simulations, how this factor varies with the metallicity and grain size is not well understood for the variety of disc conditions where dust concentration can occur. For example, while in this paper we have highlighted the case where the lack of grain growth can potentially be overcome by higher metallicities (see Figure \ref{fig:radialsmalldustprofiles}), the opposite case may also be true. If some amount of dust can reach $\mathrm{St}=1$, but at a lower abundance $Z<0.01$, dust concentration and collapse may still be possible.

Here we have only considered that only one dust size species dominates the dust-gas interaction and thus determines the local concentration of dust. A more realistic approach would be to include multiple dust sizes that are weighted towards smaller grains comparable to the ISM distribution, such as the polydisperse accretion model in \citet{Lyra2023}. One challenge in doing so is properly understanding the gas-dust backreaction in fluids with multiple dust species, which can affect the growth rate of particle instabilities and clumping \citep{Krapp2020,Zhu2021,Yang2021}. We have also only considered a disc around a solar mass star with a simple treatment of the disc temperature structure and thermodynamics. Stars with different masses will have discs of different sizes and temperatures that could change the timescales explored in this paper.

Ultimately, perhaps the most important test of this hypothesis will come from global hydrodynamic models of dusty self-gravitating discs, which may need to be divided into two epochs: the concentration and gravitational collapse of the dust and the evolution of the resulting embryos. Some existing work has already explored some aspects of either side of this process, i.e. the concentration of centimetre-sized grains in gravitoturbulent discs \citep{Boley2010} and the migration of rocky bodies in particularly massive discs \citep{Baruteau2008,Rowther2020}. In the meantime, 1D models of planet growth and migration in a viscous disc could shed some light on the viability of the solid and gas accretion explored in this paper.

\section{Conclusions}
\label{sec:conclusion}

In this paper, we have proposed a twist on the role of massive, gravitationally unstable discs in the context of planet formation. Rather than forming gas giant planets through the direct collapse of the gas disc, we assume that a quasi-steady gravitoturbulent disc can concentrate and dust particles can gravitationally collapse before the gas. This can lead to the formation of rocky bodies the size of planetary embryos that nucleate the gas disc and serve as the seeds for planet formation. The formation of a gas giant planet follows the traditional core accretion model where subsequent growth and evolution is dominated by the accretion of solids by pebble accretion in a massive disc where gas is plentiful. In this way, it is possible to explain gap and ring structures as well as the formation of giants in young discs. We summarise the important steps and timescales as follows:
\begin{itemize}
\item The growth of dust up to millimetre sizes. How long it takes for micron-sized dust of the interstellar medium to reach this size is not clearly understood as it can depend on turbulence in the disc and physical properties of the dust.
\item The concentration and gravitational collapse of millimetre dust in the dense gas structures of a marginally unstable gas disc. This process can occurs very quickly, nearly on dynamical timescale $\Omega^{-1}$ and thus we anticipate that once a disc develops gravitational instabilities with millimetre dust, planetary cores can form within a few orbits or a few thousand years depending on the radial location.
\item The seed must then accrete solids up to a mass that can begin to rapidly accrete surrounding gas. Without plentiful planetesimals, we assume that pebbles provide all the solid mass that is accreted onto these initial cores. Reaching this crossover mass for runaway gas accretion is easier in the outer disc and attainable within $10^{6}$ years for most regions of the disc.
\item Once the crossover mass is reached, the planet can grow rapidly up to a few $M_{\mathrm{Jup}}$ and takes on the order of a few thousand years to reach the thermal mass. At the thermal mass, spiral wakes excited by the planet becomes strong shocks and a gap in the disc can potentially be opened, both of which are potentially observable signatures of a planet embedded in a disc.
\end{itemize}

A few conditions must be met for this scenario to result in gas giant planets at wide separations. First, there must be some grain growth up to millimetre sizes by the time the disc becomes marginally gravitationally unstable. This is necessary not only for the formation of large seed objects but also for the efficient accretion of pebbles to grow the seed mass. Whether dust grains can grow to the necessary sizes for concentration, collapse and accretion remains uncertain and challenging both from an observational and theoretical perspective. Second, this assumes that gravitoturbulence does not last long after the formation of any planetary embryos. The strong accretion of a gravitoturbulent disc could drive rapid migration that rapidly removes any objects from their formation location before they can accrete and grow.

This provides a potential planet formation pathway that is fast in the outer disc while still maintaining the bottom up formation paradigm of the traditional core accretion scenario. This could explain the increasing number of planets suspected to exist within young discs at large orbital separations based on dust gaps in thermal emission and gas kinematic signatures.

\section*{acknowledgments}
The author thanks the anonymous referee for important comments which improved the quality of the paper. HB thanks Zhaohuan Zhu, Chao-Chin Yang, Cassandra Hall, Wenrui Xu and Aleksandra Kuznetsova for valuable comments and discussions. HB also gives special thanks Bill Watterson and his Calvin and Hobbes comic strip for inspiring the visual style of Figure 1. This research was supported in part by NASA Theoretical and Computational Astrophysics Networks (TCAN) award 80NSSC19K0639.

\section*{Software}
Matplotlib \citep{Hunter2007}, SciPy \& NumPy \citep{Virtanen2020,vanderWalt2011}, and IPython \citep{Perez2007}

\section*{Data Availability}

The data underlying this article will be shared on reasonable request to the corresponding author.



\bibliographystyle{mnras}
\bibliography{library} 








\bsp	
\label{lastpage}
\end{document}